\documentclass[12pt,preprint]{aastex}

\newcommand{\et}{et al.}

\newcommand{\ha}{H$\alpha$}
\newcommand{\solar}{\ifmmode_{\sun}\;\else$_{\sun}\;$\fi}

\newcommand{\xxx}{\enspace\enspace\enspace}

\begin{document}

\title{{\it GALEX} Ultraviolet Imaging of Dwarf Galaxies and 
Star Formation
Rates\footnote{Based on observations made with the NASA {\it Galaxy Evolution Explorer} ({\it GALEX}).
{\it GALEX} is operated for NASA by the California Institute of Technology under
NASA contract NAS5-98034.}}

\author{Deidre A. Hunter}
\affil{Lowell Observatory, 1400 West Mars Hill Road, Flagstaff, Arizona
86001 USA}
\email{dah@lowell.edu}

\author{Bruce G. Elmegreen}
\affil{IBM T. J. Watson Research Center, PO Box 218, Yorktown Heights,
New York 10598 USA}
\email{bge@watson.ibm.com}

\and

\author{Bonnie C. Ludka\footnote{Current address: Scripps Institution of Oceanography,
University of California, San Diego}}
\affil{Lowell Observatory, 1400 West Mars Hill Road, Flagstaff, Arizona 86001 USA}
\email{ludkabc@gmail.com}

\begin{abstract}
We present ultraviolet integrated and azimuthally-averaged surface
photometric properties of a sample of 44 dIm, BCD, and Sm galaxies
measured from archival NUV and FUV images obtained with {\it
GALEX}. We compare the UV to \ha\ and $V$-band properties and convert
FUV, \ha, and $V$-band luminosities into star formation rates (SFRs).
We also model the star formation history from colors and compare the
integrated SFRs and SFR profiles with radius for these methods. In most
galaxies, the UV photometry extends beyond \ha\ in radius, providing a
better measure of the star formation activity in the outer disks. The
H$\alpha$ appears to be lacking in the outer disk because of
faintness in low density gas. The FUV and $V$-band profiles are
continuous with radius, although they sometimes have a kink from a
double exponential disk. There is no obvious difference in star
formation properties between the inner and outer disks. No disk edges
have been observed, even to stellar surface densities as low as
$0.1\;M_\odot$ pc$^{-2}$ and star formation rates as low as
$10^{-4}\;M_\odot$ yr$^{-1}$ kpc$^{-2}$. Galaxies with low HI to
luminosity ratios have relatively low FUV compared to $V$-band emission
in the outer parts, suggesting a cessation of star formation there.
Galaxies with relatively high HI apparently have fluctuating star
formation with a Gyr timescale.
\end{abstract}


\section{Introduction} \label{sec-intro}

Outer edges of dwarf irregular (dIm) galaxies present an extreme environment for star formation.
Dwarf galaxies already challenge models of star formation because of their low gas densities
even in the central regions
(Hunter \& Plummer 1996, Meurer \et\ 1996, van Zee \et\ 1997, Hunter \et\ 1998,
Rafikov 2001).
Outer parts of dwarfs, where the gas density is even lower, therefore,
present a particularly difficult test of our understanding of the cloud/star formation process.
Yet, from broad-band images we see that stars have formed in the outer parts to very
low surface brightness levels (for example, D.\, Hunter \et\ 2009, in preparation).

We are using UV images obtained with the {\it Galaxy Evolution Explorer}
satellite ({\it GALEX}; Martin \et\ 05)
to trace and characterize star
formation from the centers out
into the outer disks of dwarf galaxies where \ha\ may not be an effective tracer of recent star
formation (see, for example, Thilker \et\ 2005, 2007a; Boissier \et\ 2007).
Fortunately, the UV---ultraviolet light coming directly from OB stars---can easily
trace young stars in the outer galaxy, and
the UV has the advantage in a patchy star-forming environment of integrating over a
longer timescale ($\leq$100--200 Myrs) than does \ha\ ($\leq$10 Myrs).

In this paper we compare the UV surface brightness profiles to those of
\ha\ and to those of the older stars as seen in broad-band images.
We compute integrated and azimuthally-averaged profiles of several measures
of star formation rates (SFRs): derived from FUV, \ha, and $V$-band luminosities and
model stellar population fits to broad-band UV, optical, and near-infrared colors.
In a companion paper, we examine characteristics of individual star-forming regions identified
on the {\it GALEX} images and compare those in the outer disk with those in the inner disk
(Melena \et\ 2009).
The galaxies discussed in these two studies are also part of a larger, multi-wavelength
survey of dIm, Blue Compact Dwarf (BCD), and Sm galaxies
that has been assembled in order to examine the drivers
of star formation in tiny galaxies (Hunter \& Elmegreen 2004, 2006).

\section{The Data} \label{sec-data}

The galaxies in this investigation are a sub-sample of a large
multi-wavelength survey of 136 relatively normal nearby galaxies
without spiral arms (94 dIm, 24 Blue Compact Dwarfs, and 18 Sm
galaxies; Hunter \& Elmegreen 2004, 2006). That survey is
representative of the range in galactic parameters exhibited by dwarfs,
and is aimed at understanding star formation processes in dwarf
galaxies. The $UBVJHK$ and \ha\ data obtained as part of that survey
are described by Hunter and Elmegreen.

\begin{figure}
\epsscale{1.0}
\plotone{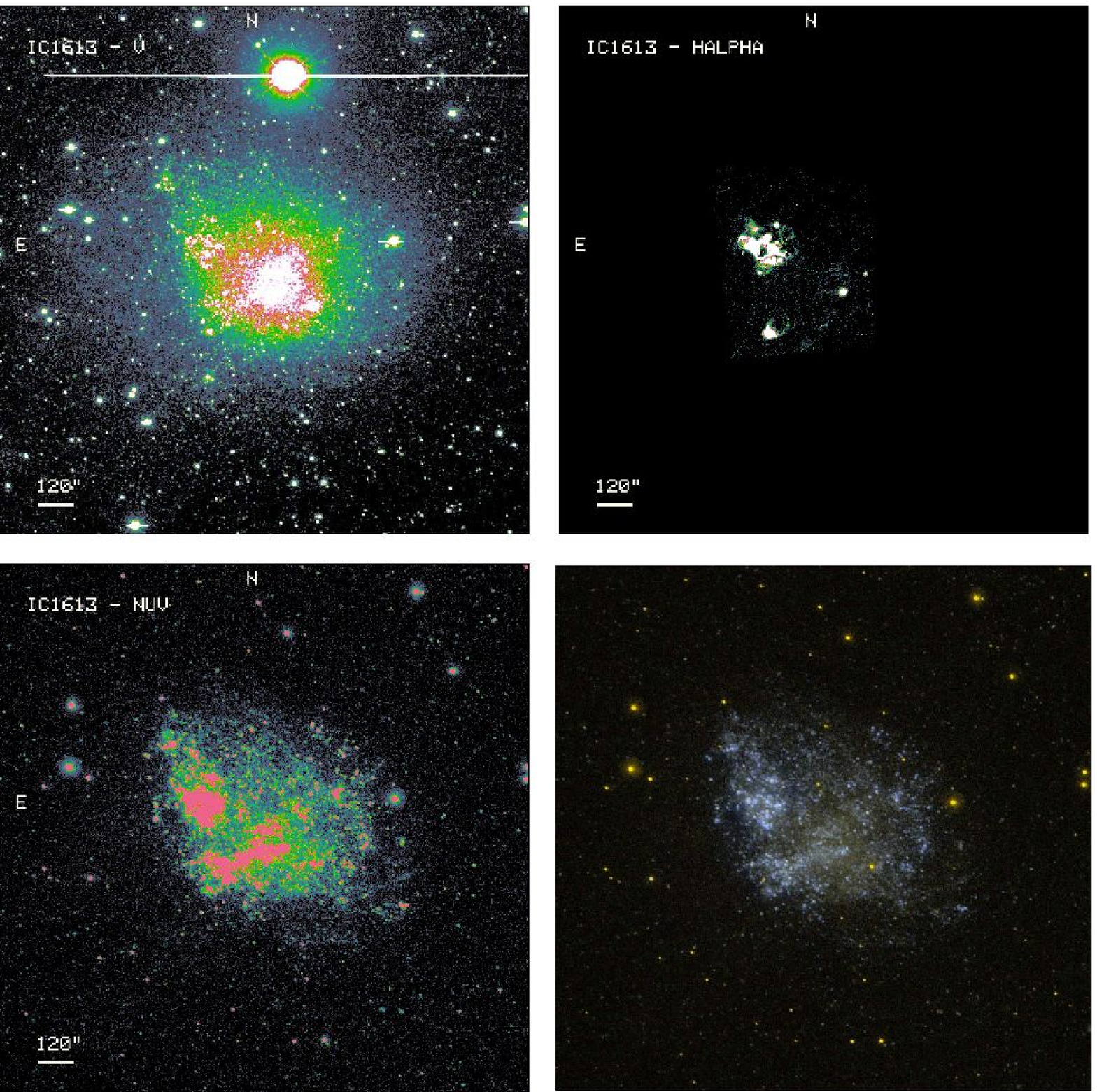}
\caption{Images of IC 1613 as an example of the data. $V$, \protect\ha, and {\it GALEX} NUV
images are shown in false-color and with the same field of view and pixel scale.
The image in the lower right is from the {\it GALEX} pipeline and is a combination
of the FUV and NUV images. It has nearly the same field of view as the other images.
The $V$ and \protect\ha\ images are from Hunter \& Elmegreen (2004, 2006).
\label{fig-ic1613}}
\end{figure}

To extend this survey into the UV, we obtained images from the {\it
GALEX} archives for 44 of our 136 dwarf galaxies, including
29 dIm, 8 BCDs, and 7 Sm galaxies.
The galaxies are listed in Table \ref{tab-genobs} along with
the exposure times and tile name of the {\it GALEX} images.
{\it GALEX} images simultaneously in two channels: FUV with a
bandpass of 1350--1750 \AA, an effective wavelength of 1516 \AA, and
a resolution of 4.0\arcsec\ and NUV with a bandpass
of 1750--2800 \AA, an effective wavelength of 2267 \AA,
and a resolution of 5.6\arcsec.
The images were processed through
the {\it GALEX} pipeline and were retrieved as final intensity maps with a 1.5\arcsec\ pixel scale.
The {\it GALEX} field of view is a circle with 1.2\arcdeg\ diameter, and we have extracted
a portion around our target galaxies.
Figure \ref{fig-ic1613} shows $V$, \ha, and UV images of IC 1613 as an example of the data.
The \ha\ and $V$-band images are described by
 (Hunter \& Elmegreen 2004, 2006).
In Table \ref{tab-opt} we collect integrated optical $UBV$, near-IR $JHK$, and \ha\
photometry of our sample from Hunter and Elmegreen.

We subtracted or masked foreground stars and background galaxies, and
removed sky from the {\it GALEX} NUV and FUV images. In some cases
the sky was a constant determined from regions around the galaxy, but
in other cases the sky was determined from a low-order, two-dimensional
fit to the surroundings. We geometrically transformed the UV images to
match the orientation and scale of our \ha\  and $V$-band images. We
measured surface photometry in the NUV and FUV images in fixed
ellipses of increasing semi-major axis length, using the center,
position angle, ellipticity, and step size that were used to do surface
photometry on the optical images Hunter \& Elmegreen (2006).

We corrected the UV photometry for extinction using the same total reddening E($B-V$)$_t$ as
was used for correcting the $V$-band photometry in the larger survey:
foreground E($B-V$)$_f$ from
Burstein \& Heiles (1984) given in Table \ref{tab-genobs} plus
0.05 mag for internal reddening of the stars. The \ha\ photometry was corrected with 0.1 mag for
the internal reddening in HII regions.
This  level of internal reddening is consistent with measurements
of the Balmer decrement in HII regions in a sample of 39 dIm galaxies (Hunter \& Hoffman 1999).
There the average reddening in HII regions is 0.1, and we have taken half this to represent
the stars outside of HII regions.

We combined the E($B-V$)$_t$ with the extinction law
of Cardelli \et\ (1989) to produce the extinction
$A_{FUV}=8.24{\rm E}(B-V)_t$, and, 
interpolating over the 2175 \AA\ bump, $A_{NUV}=7.39{\rm E}(B-V)_t$. 
However, the FUV extinction law varies from galaxy to galaxy, and
the NUV filter straddles the 2175 \AA\ extinction feature, which also varies not only from galaxy
to galaxy but from place to place within galaxies (for example, Gordon \et\ 2003).
Star formation activity appears to play as important a role, perhaps even more important,
than metallicity in determining the UV extinction curve and the strength of the 2175 \AA\ bump.
In the metal-poor SMC, for example, Gordon \& Clayton (1998) have shown that lines of sight through
the actively star forming bar have UV extinction curves that are linear with 1/$\lambda$, rising more 
steeply into the FUV than in the Milky Way, and no 2175 \AA\
bump is present, while a sight line through the quiescent wing has an extinction curve that is 
similar to that in the Milky Way with a shallower slope into the FUV and a 2175 \AA\ bump. 
Similarly in the LMC, regions of intense star formation have stronger extinction into the FUV and weaker 
2175 \AA\ than the average LMC extinction curve (see, for example, Gordon \et\ 2003).
Furthermore, the reddening parameter $R_V=A_V/{\rm E}(B-V)$ varies by 25\% within the LMC, for example.
In our dwarf galaxies, the level of star formation activity is highly variable from galaxy to galaxy
and from place to place within a galaxy. 
To complicate things further, the reddening we need to correct is the result of two different
reddening laws: that due to the Milky Way and that due to the dwarf galaxy under study.
Wyder \et\ (2007) have dealt with this for a large sample of galaxies by
adopting the Cardelli \et\ extinction law for the Milky Way 
and determining the best average reddening
from convolving theoretical galaxy spectral energy distributions (SEDs) 
with the {\it GALEX} filter transmission curves.
In the same manner, we have examined the results of convolving different SMC, LMC, and Milky Way
extinction curves (Cardelli \et 1989, Gordon  \et\ 2003)
with constant star formation rate galaxy SEDs constructed from the Bruzual \& Charlot
(2003) stellar population library and with the {\it GALEX} filter transmission curves.
The $A_\lambda/{\rm E}(B-V)$ vary by 1.5 magnitudes
depending on the extinction curve.
Considering the variety of environments represented by our sample and the fact that 
extinction is minimal except for a few galaxies with higher than average foreground
extinction,
we decided to adopt the Wyder \et\ extinctions:
$A_{FUV}=8.24{\rm E}(B-V)_t$ and $A_{NUV}=8.2{\rm E}(B-V)_t$.
For most of our galaxies, the variations represented by alternate extinction laws result
in differences of order 0.05 mag.

We used a single extinction correction for all radii within each galaxy.
This is justified by the lack of metallicity gradients in dwarfs and the correlation
between metallicity and attenuation seen in spirals (Boissier \et\ 2007).
On the other hand, nothing is known about the outer disks of dwarfs, beyond
where \ha\ emission is no longer detected, and there is some evidence for
large quantities of cold dust out there (Galliano \et\ 2003).
However, Boissier \et, based on {\it IRAS} observations, suggest that there is no attenuation
in the outer disks of two dwarfs (IC1613, WLM), but uncertainties are large due to low {\it IRAS}
fluxes.

Photometry in the largest ellipse yields integrated parameters, and differences between the ellipses
divided by the area of the annulus gives azimuthally-averaged surface photometry.
We fit the surface photometry with an exponential disk: $\mu^{NUV}(R) = \mu_0^{NUV} +
1.086R/R_D^{NUV}$, where $R_D^{NUV}$ is the disk scale length measured in the
NUV passband. The same quantity measured in the $V$ filter is denoted $R_D^V$.
The fit was made with uniform
weighting of each radial data point. Some galaxy profiles were better fit with two lines,
as was sometimes also the case in the $V$-band, and some profiles were not well fit
by a straight line at all.

\section{Integrated Photometry} \label{sec-int}

In Table \ref{tab-phot} we give the integrated UV photometry, including the NUV absolute magnitude
$M_{NUV}$, the luminosity $L_{NUV}$, the color $(FUV-NUV)_0$, and the color $(NUV-V)_0$.
Note that $M_{NUV}$ is an AB magnitude, while magnitudes from the $UBVJHK$ passbands
are Vega magnitudes.
All quantities are corrected for reddening.
We also give the ratio $L_{H\alpha}/L_{NUV}$.
The $V$-band and \ha\ surface photometry are from Hunter \& Elmegreen (2004, 2006).

\begin{figure}
\epsscale{1.0}
\plotone{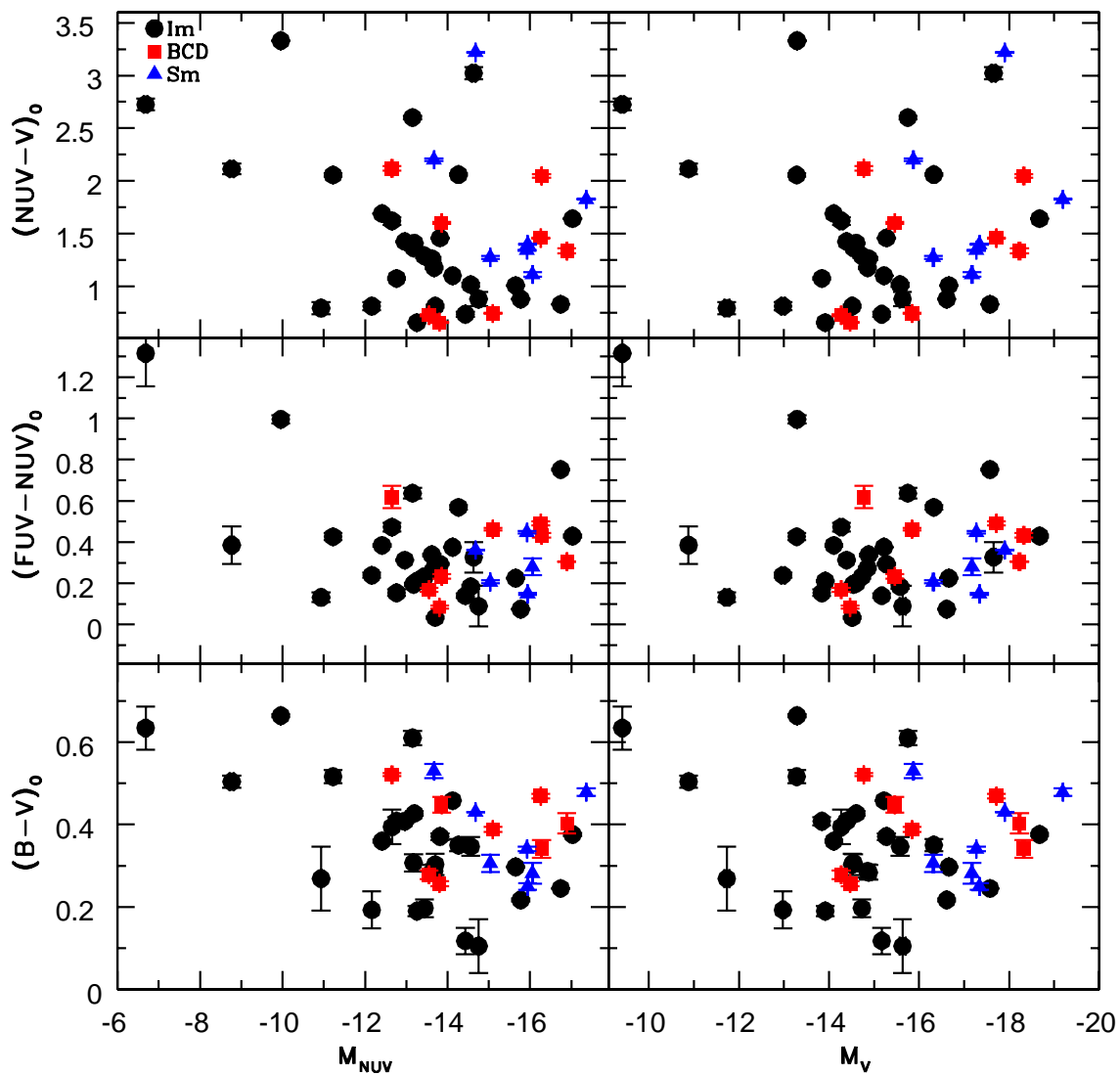}
\caption{Integrated colors plotted against the absolute NUV AB magnitude ({\it left})
and absolute $V$ magnitude ({\it right}).
Colors are corrected for reddening.
The dIm sample is shown as {\it black filled circles}, BCDs as {\it red filled squares},
and Sm galaxies as {\it blue filled triangles}.
There is a rough trend of redder optical color for lower UV luminosity.
\label{fig-mnuv}}
\end{figure}

\begin{figure}
\epsscale{1.0}
\plotone{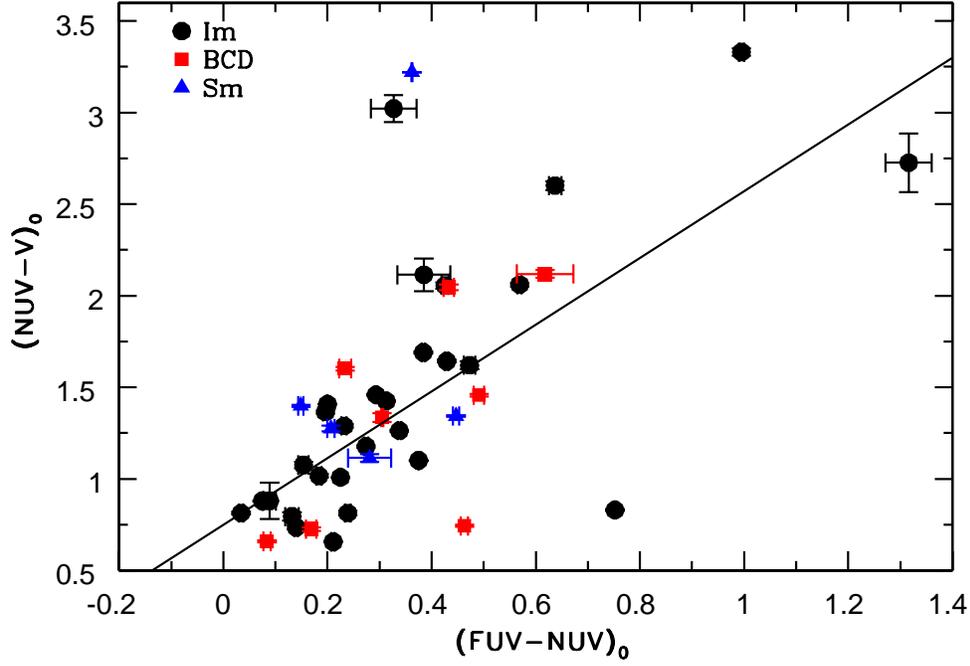}
\caption{Integrated UV-optical color-color plot.
Colors are corrected for reddening.
The dIm sample is shown as {\it black circles}, BCDs as {\it red squares},
and Sm galaxies as {\it blue triangles}.
The solid black line is a fit to all of the points except the 4 galaxies with
the highest $(NUV-V)_0$: \protect\\
$(NUV-V)_0 = (0.75 \pm 0.12) + (1.82\pm0.27)\times (FUV-NUV)_0$.
The two colors track each other.
\label{fig-col}}
\end{figure}

In Figure \ref{fig-mnuv} we plot $(FUV-NUV)_0$ and $(NUV-V)_0$, as well
as $(B-V)_0$, against $M_{NUV}$ and $M_V$. There is a rough trend of redder optical color
for lower luminosity in the UV. However, the scatter is large and the low luminosity end
of the distribution is held down by just a few points.
We do not see the trend of bluer UV-optical colors with lower luminosity
found in studies of large samples of more massive galaxies (e.g. Wyder \et\ 2007).
In an integrated color-color plot (Figure \ref{fig-col})
we do see a correlation: As a galaxy becomes redder
in $(FUV-NUV)_0$ it also becomes redder in $(NUV-V)_0$.

Most of the galaxies have an integrated $(FUV-NUV)_0$ color between 0 and 0.5 magnitudes.
According to Thilker \et\ (2005), a constant star formation rate stellar population
would have an $(FUV-NUV)_0$ color of order $-0.1$ to 0 magnitudes.
So the dwarfs are generally redder than this model.
However, the colors of most of the dwarfs are comparable to those of extinction-corrected
colors of star-forming galaxies in the Goldmine sample as presented by Boissier \et\ (2008).
A few of the dwarfs have redder colors than this sample, and are comparable to the
colors of low surface brightness galaxies of Boissier \et, which they interpret as most likely
due to a variable star formation history.
The BCDs in our sample cover the same range of integrated colors as the dIm galaxies.
However, most of these galaxies have $(FUV-NUV)_0$ color gradients and are
bluer in their centers, consistent with the central concentration of star formation that
characterizes the BCD class of galaxies.

\section{Surface Brightness Profiles} \label{sec-sb}

\begin{figure}
\epsscale{0.9} \plotone{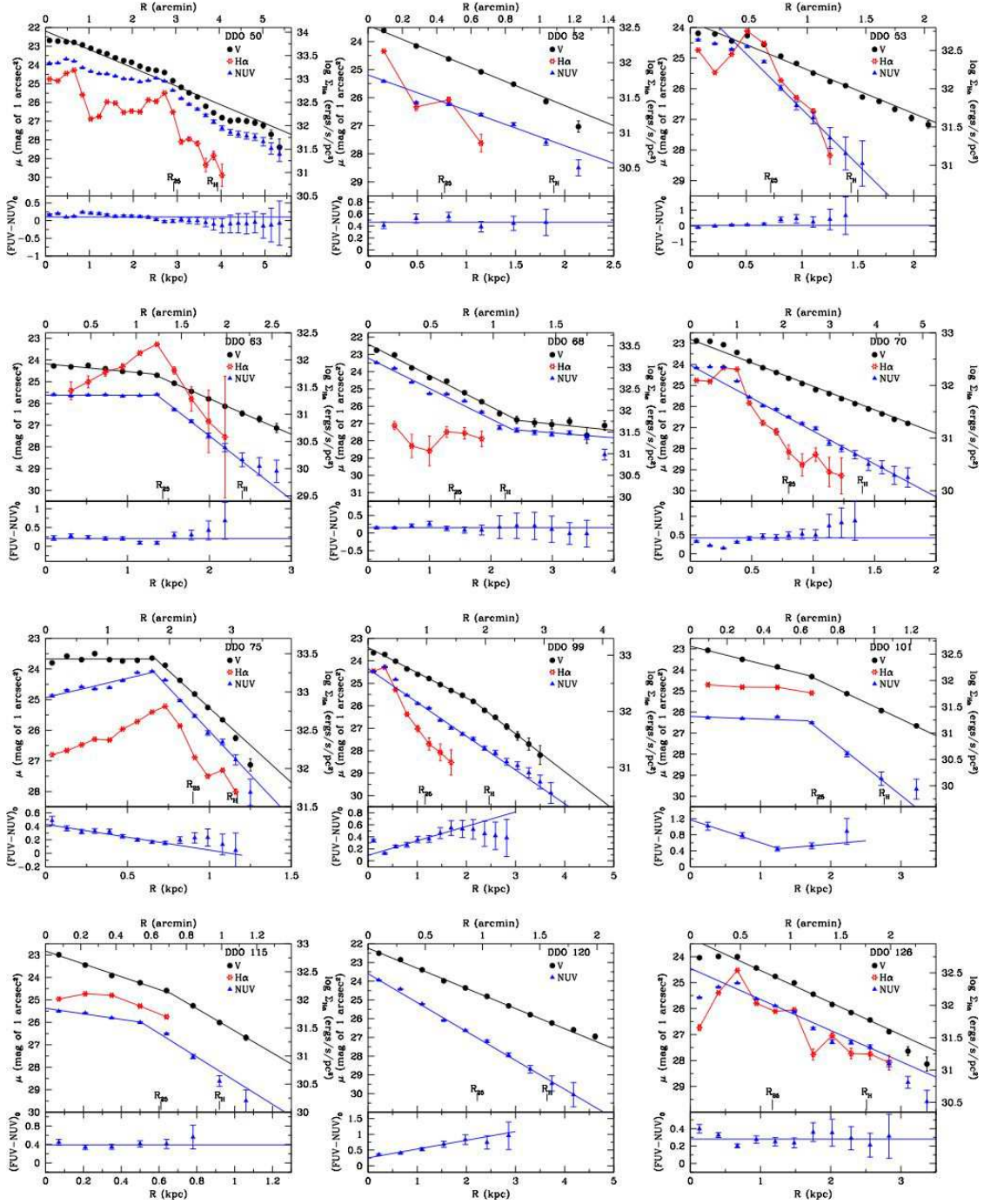}
\caption{Azimuthally-averaged surface brightness profiles in $V$ ({\it
black}), \protect\ha\ ({\it red}), {\it GALEX} NUV ({\it blue}), and
$(FUV-NUV)_0$ ({\it bottom panels}). For NGC 3109, there is no $V$-band
image and a profile is shown instead for an \protect\ha\ off-band
filter centered at 6440 \AA\ with a FWHM of 95 \AA. The radii $R_{25}$
and $R_H$ are marked. $R_{25}$ is the radius at which the $B$-band
surface brightness drops to 25 mag arcsec$^{-2}$. $R_H$ is the Holmberg
radius, the radius at which the $B$-band surface brightness drops to
about 26.7 mag arcsec$^{-2}$. \label{fig-profiles}}
\end{figure}

The surface brightness profiles are shown in Figure \ref{fig-profiles}.
There we compare surface photometry in NUV, $V$, and \ha. The UV
profiles extend beyond the \ha\ profiles by at least two annuli in 66\%
(19 of 29) of the dIm, 75\% (6 of 8) of the BCDs, and 71\% (5 of 7) of
the Sm galaxies. The uncertainty in these fractions
(taken from the square root of the number of galaxies)
in the BCD and Sm samples
is of order 35\%, while the uncertainty for the dIm galaxies is 19\%.
Four dIm galaxies without \ha\ emission have UV emission; all of the
galaxies have some UV emission.

We have fit the NUV surface photometry with exponential disk profiles,
where appropriate, and the central surface brightness and disk scale lengths are
given in Table \ref{tab-struc}.
There are two sets of entries in this table---measures
of an inner profile and measures of an outer profile. Those galaxies whose profiles
are best fit with a double exponential have two sets of numbers, while those that
are fit just fine with a single exponential have entries only under the heading of
``inner'' profile. Of the 29 dIm galaxies, 8 have clear double exponential profiles
in NUV; in 7 of these the outer exponential drops more steeply than the inner,
and in one the outer exponential is shallower.
Three other galaxies show breaks in $V$, but not in the UV.
Of the 8 BCDs, 5 have NUV double
exponentials, all of which drop more shallowly in the outer disk.
And, 4 of the 7 Sms have profiles with breaks, all of which drop more steeply in
the outer disk.

One galaxy, Haro 3, deserves individual mention.  The UV image of Haro
3 was taken from archive combined images of unusual length. The NUV
exposure altogether was 8.4 hours and the FUV exposure was 3.5 hours
rather than the typical exposure time of order half of an hour. Because
of the unprecedented (for extragalactic observations) exposure times,
the signal-to-noise is superb and we are able to trace NUV to a
surface brightness level of 29 magnitudes of 1 arcsec$^2$. This is
about twice as far in radius as our $V$-band and \ha\ photometry go. In
Figure \ref{fig-profiles} one can see that at the radius where $V$ and
\ha\ end, the NUV surface photometry changes slope, becoming
shallower. The break occurs at $2.3R_D^{NUV}$, where $R_D^{NUV}$ here
is the disk scale length determined from the {\it outer} part of the
profile. This type of double exponential is typical of many BCDs in
optical and near-IR passbands, and the shallower outer profile is
generally accepted as representing the underlying stellar
population while the steeper inner profile is dominated by the intense
recent star formation that is concentrated in the galaxy center (see,
for example, Papaderos \et\ 1996, Cairos \et\ 2001, Noeske \et\ 2003,
Hunter \& Elmegreen 2006).
Consistent with this picture, we see in Figure \ref{fig-profiles} that
$(FUV-NUV)_0$ is bluer in the inner part of the galaxy than in the outer parts,
although the upturn to redder $(FUV-NUV)_0$ begins much closer to the center than the
break in $\mu_{NUV}$.

In Table \ref{tab-struc} we give the $V$-band disk scale lengths
$R_D^V$ from Hunter \& Elmegreen (2006) for comparison, and in Figure
\ref{fig-rd} we plot $R_D^{NUV}$ against $R_D^V$. Symbol types indicate
whether these scale lengths refer to the inner or outer disks. We see
that the disk scale lengths in the two passbands are similar in most of
the galaxies. Small galaxies, mostly Im and BCD, have slightly larger
$R_D^V$ than $R_D^{NUV}$, which means that their NUV emission and hence
recent star formation is more centrally
concentrated. Large galaxies, usually Sm, have slightly smaller $R_D^V$
than $R_D^{NUV}$, which means their most recent star formation includes 
the outer disk.

In Figure \ref{fig-profiles} we include azimuthally-averaged
$(FUV-NUV)_0$ as a function of radius. Of the 29 dIm galaxies, 18
(62\%) have constant colors as far as we measure both UV passbands,
similar to low surface brightness galaxies (Boissier \et\ 2008).
However, 7 (24\%) dIm galaxies get redder with radius, one galaxy gets
bluer, and 3 have color profiles that are too complex to put into one
of these categories. Of the 8 BCDs, on the other hand, 6 (75\%) get
redder with radius, as one would expect for centrally concentrated star
formation, while two have constant colors. Interestingly, of the 5 Sm
galaxies measured in both UV passbands, 4 have constant $(FUV-NUV)_0$
colors with radius and one profile gets slightly bluer with radius. The
average $(FUV-NUV)_0$ colors, if constant, or color gradients are given
in Table \ref{tab-colorgrad}.

The double exponential profiles observed in $\mu_V$ and $\mu_{\rm NUV}$
in some of our dwarfs
are also seen in some spiral disks. Roskar \et\ (2008) use simulations
to argue that the break is a combination of reaching a star formation
threshold in the outer disk and scattering of stars outward from the
inner disk by spiral arms. This predicts that the stellar population in
the outer disk is older than that in the inner disk since many of the
stars did not form there. Therefore, we expect a redder stellar
population in the outer disk. This is not what we see in dIm galaxies.
Most often colors in dIm galaxies are constant into the outer disk from
the ultraviolet to the near-IR, and there is no correlation between
dwarf galaxies that do show a color gradient and the presence of double
exponential profiles. The surface brightness profile breaks, when
present, are seen in all passbands from the UV to the near-IR (c.f.,
Hunter \& Elmegreen 2006, Hunter \et\  2006). Furthermore, dwarf
irregulars do not have strong spiral arms, so the scattering mechanism
proposed by Roskar et al.\ cannot operate in these systems.
Stellar scattering in bar-like potentials might be important for some 
dwarfs, however.

\section{Integrated Star Formation Rates} \label{sec-intsfr}

\subsection{FUV, \ha, and $V$-band luminosities}

We have converted the integrated photometry and azimuthally-averaged surface photometry
in the FUV, \ha, and  $V$-band into SFRs. For the FUV, we use the formula
from Kennicutt (1998) modified for the sub-solar metallicities appropriate for dwarf galaxies.
Kennicutt's original formula is for $Z$\solar. We used the stellar population evolution
models of STARBURST99 (Leitherer \et\ 1999) to determine the scaling from solar
metallicity to sub-solar. Their Figure 54 plots the 
monochromatic luminosity at 1500 \AA\ as a function of time for a constant star formation
rate stellar population with a Salpeter (1955) stellar 
initial mass function (IMF)---the function describing the
number of stars of a given mass,
and stellar masses 0.1--100 M\solar.
The effective wavelength of the FUV filter is 1516 \AA, 
so using the luminosity at 1500 \AA\ to examine the effect of metallicity is reasonable.
The ratio of the luminosity for
$Z=0.008$ to that for $Z$\solar\ is 1.10 at 100 Myrs, and we have applied the
inverse of this factor to Kennicutt's proportionality constant to obtain the following:
$$SFR_{\rm FUV} (M\solar\ {\rm yr^{-1}}) = 1.27\times10^{-28} L_{FUV} {\rm (ergs ~s^{-1} Hz^{-1})}.$$
The proportionality constant used here
is 18\% higher than the empirically determined relationship determined by Salim \et\ (2007)
for sub-solar (roughly 0.8$Z$\solar) metallicities 
in star-forming galaxies with a variety of star formation histories.
On the other hand, our proportionality is 26\% lower than that used by Thilker \et\ (2007b), taken from 
Iglesias-P\'aramo \et\ (2006) who used STARBURST99 models at solar metallicity.

For consistency in comparing SFRs, we also use Kennicutt's (1998) formula for determining the
SFR from \ha, modified again for non-solar metallicity.
We scaled Kennicutt's formula by the
number of photons with wavelengths below 912 \AA\ available for
ionizing the gas using STARBURST99 (Leitherer \et\ 1999). 
The ratio of the number of ionizing photons at
$Z=0.008$ to that at $Z$\solar\ is 1.15, and we have applied the
inverse of this factor to Kennicutt's proportionality constant to obtain the following:
$$SFR_{\rm H\alpha} (M\solar\ {\rm yr^{-1}}) = 6.9\times10^{-42} L_{H\alpha} {\rm (ergs ~s^{-1})}.$$
Note that this formula produces a SFR that is 1.1 times higher than
that which we derived and used for our larger optical survey (see
Appendix B of Hunter \& Elmegreen 2004). 
Thilker \et\ (2007b) used a formula for solar metallicity, taken from Hirashita \et\ (2003) and
based on STARBURST99, that is the same as Kennicutt's.

To determine the SFR averaged over a longer time scale, we use the $V$-band luminosity to
determine the mass in stars and assume a timescale of 12 Gyr for the formation of that stellar mass.
We use a stellar $M/L_V$ ratio from Bell \& de Jong (2001)
that is a function of $(B-V)_0$ color: 
$\log(M/L_V) = -0.37 + 1.14 (B-V)_0.$
If there is no $(B-V)_0$ measurement or
the uncertainty in the color is $>0.1$ magnitude, we assumed $M/L_V=1.07$
(appropriate for $(B-V)_0=0.35$). 
The Bell and de Jong relationship between mass-to-light ratio and color 
that we use is appropriate for a Salpeter (1955) stellar IMF and metallicity 
$Z=0.008$ and is derived using the Bruzual \& Charlot (2003) stellar population
models.
The SFR determined from the $V$-band is
the SFR necessary to produce the original mass in stars.
The $M/L_V$ formula gives us the mass currently in stars, and we 
correct this for the mass recycled back into the interstellar medium.
We use a factor of 2 from Brinchmann \et\ (2004) for this correction. 
The final formula for converting $V$-band to a SFR is as follows:
$$\log SFR_{\rm V} (M\solar\ {\rm yr^{-1}}) = -0.4 M_V + \log (M/L_V) -7.842.$$

The integrated SFRs are given in Table \ref{tab-sfr}, and plotted
against each other in Figure \ref{fig-intsfr}. The integrated $SFR_{\rm
H\alpha}$ tends to be lower than $SFR_{\rm FUV}$ for some of the lower
SFR dIm galaxies by a factor of 2.5 on average, but matches well for
the lower SFR BCD and Sm galaxy sub-samples. At the high SFR end,
$SFR_{\rm H\alpha}$ is a little high by a factor of up to 3 compared to
$SFR_{\rm FUV}$ for several dIm, BCD, and Sm galaxies. On the other
hand, the SFR determined from $M_V$, $SFR_{\rm V}$, tends to be comparable
on average to the current integrated $SFR_{\rm FUV}$
for all dwarf types except the very lowest SFR systems. 
There is, however, scatter around the relationship up to a factor of 3 for most of
the galaxies.
The low SFR
systems, on the other hand, have higher $SFR_{\rm V}$ than current integrated $SFR_{\rm FUV}$. 
We consider the
differences between these SFR measures in the following subsections.

\subsubsection{Comparison of FUV and H$\alpha$}\label{fuv-ha}

\subsubsubsection{Extinction, Calibration, and Age Effects}

$SFR_{\rm H\alpha}$ and $SFR_{\rm FUV}$ are both measures of recent star formation
activity. Although they are sensitive to somewhat different time scales, we do expect these
two SFR indicators to be fairly close. Here we consider possible reasons for the
differences that we see.

Our first consideration is whether extinction errors contribute to the
discrepancy between $SFR_{\rm H\alpha}$ and $SFR_{\rm FUV}$. The cause
of the discrepancy is unlikely to be this simple given that our
integrated $SFR_{\rm H\alpha}$ are a little high at the high SFR end
while being too low at lower SFRs, but it is worth exploring. Treyer
\et\ (2007) have analyzed Sloan Digital Sky Survey and {\it GALEX} data
on a large sample of galaxies to determine integrated FUV extinctions
and SFRs. They assume that extinctions determined from the Balmer
decrement and SFRs determined from emission lines are most accurate.
They then derive extinction formulae to make the UV-based SFRs come
into agreement with the \ha-based SFRs. The extinctions that their
formulae imply for the dwarf galaxies are much higher than what we use
here. On the other hand, dust attenuation as a function of galactic
mass determined from models give \ha\ attenuations that are very close
to those that we are using (Brinchmann \et\ 2004). Furthermore, in both
studies extinctions determined from the Balmer decrement are considered
to be the best, and our average internal E$(B-V)_i$ for \ha\ emission
is derived from emission-line spectroscopy although a single average
value is applied to all galaxies (Hunter \& Elmegreen 2004). In any
event, increasing the extinction in the UV will not make the
differences that we see go away since the UV-based SFRs are high
compared to the \ha-based SFRs. We would need instead to raise the
extinction selectively to \ha\ by 0.95 magnitude (0.37 magnitude in
E($B-V$)$_0$), but only in the lower SFR systems. Such high extinction
values would be inconsistent with average Balmer decrements measured in
dwarf galaxy HII regions.

Perhaps the difference between integrated $SFR_{\rm H\alpha}$ and
$SFR_{\rm FUV}$ is the result of an incorrect scaling coefficient for
converting \ha\ or FUV luminosity relative to each other.  
The Brinchmann \et\
(2004) models suggest that Kennicutt's (1998) scaling factor for
converting $L_{\rm H\alpha}$ to $SFR_{\rm H\alpha}$ is a function of
galaxy stellar mass. For the masses of our galaxies, a factor that is
0.7--0.8 times Kennicutt's scaling factor is suggested on average,
similar to Salim \et's proportionality constant, which is 0.77 times
Kennicutt's original value, for converting FUV to $SFR_{\rm FUV}$.
For $SFR_{\rm H\alpha}$ we are scaling the original by 0.81 in order to
account for the lower metallicity. Thus, we are using a formula that
is similar to what others have derived for similar galaxies, although there are 
certainly uncertainties.
Furthermore, there is no correlation between our ratio 
$SFR_{\rm FUV}$/$SFR_{\rm H\alpha}$ and the oxygen abundance of the galaxy.

Could the difference between integrated $SFR_{\rm H\alpha}$ and
$SFR_{\rm FUV}$ be due to a larger contribution from older stars in
galaxies where the SFR is lower? If that were the case, we would expect
to see a correlation between the UV color $(FUV-NUV)_0$ and the
$SFR_{\rm FUV}$. That is, as the SFR goes down, $(FUV-NUV)_0$ should
get redder ($FUV-NUV$ is a poor indicator of dust
attenuation [Buat \et\ 2005]). Figure \ref{fig-sfrcolor} is a plot of
$(FUV-NUV)_0$ and $(NUV-V)_0$ against our three luminosity-based SFR
measures. There is not much of a trend of either color with any SFR. In spiral
galaxies, FUV is considered a good measure of star formation activity
for $FUV-NUV<1$ (Boissier \et\ 2007), and that is likely to be the case
in dIm galaxies as well.

\subsubsubsection{IMF Effects}

These considerations suggest that variable extinction, metallicity, and
star formation history could contribute small differences between
$SFR_{\rm H\alpha}$ and $SFR_{\rm FUV}$, but they are not likely to
explain the full extent of the differences that we see. Another
interesting possibility is that the difference between $SFR_{\rm
H\alpha}$ and $SFR_{\rm FUV}$ is the result of variations in the
stellar IMF. Meurer \et\ (2009) examined integrated $SFR_{\rm H\alpha}$
and $SFR_{\rm FUV}$ for nearby galaxies and found a relationship
between this ratio and the surface brightness in \ha. The relationship
is such that galaxies with lower \ha\ surface brightnesses have higher
$SFR_{\rm FUV}/SFR_{\rm H\alpha}$ ratios. They suggest that this ratio
is also higher than predicted by stellar population evolution models
with constant SFRs, or by evolutionary histories with bursts or gasps
in the SFRs. They concluded that the relationship is due to variations
in the stellar upper mass limit or the slope of the stellar IMF.

We reproduce Meurer \et's  (2009) relationship in the top panel of Figure
\ref{fig-imf}. For H$\alpha$ surface brightness we use our integrated
H$\alpha$ SFR and divide by the area over which H$\alpha$ was
integrated. There is a clear correlation spanning 4 orders of magnitude
in $SFR_{\rm H\alpha}$ and one order of magnitude in $SFR_{\rm
FUV}/SFR_{\rm H\alpha}$.  Lower H$\alpha$ surface brightness
corresponds to decreasing H$\alpha$ flux relative to FUV flux for all
galaxy types in our survey.  A power law fits this relationship well:
$SFR_{\rm FUV}/SFR_{\rm H\alpha}\propto SFR_{\rm
H\alpha}^{-0.59\pm0.07}$. This result is similar to that in Meurer et
al., which is $SFR_{\rm FUV}/SFR_{\rm H\alpha}\propto SFR_{\rm
H\alpha}^{-0.43\pm0.03}$.  The bottom panel of Figure \ref{fig-imf}
shows no relationship between galactic $M_V$ and $SFR_{\rm
FUV}/SFR_{\rm H\alpha}$, so the $SFR_{\rm H\alpha}$ correlation is only
with surface brightness, not galaxy size. One can therefore imagine,
following Meurer, et al., that surface brightness affects the IMF by
producing proportionally more massive stars at higher interstellar
pressures.

We also considered the IMF-dependent conversion from \ha\ luminosity to
SFR proposed by Pflamm-Altenburg, Weidner, \& Kroupa (2007, hereafter
PWK). They suggest that the maximum stellar mass in a star cluster is
determined by the total cluster mass, and that the maximum cluster mass
in turn depends on the total SFR of the system. Galaxy-wide IMFs should
then be steeper than cluster IMFs, and they should be a function of the
integrated SFR of the galaxy. As a result, the SFR should be higher for
a given H$\alpha$ flux than it is with the standard IMF.

To compare our H$\alpha$-based SFRs in Table \ref{tab-sfr} with the PWK
rates, we computed new SFRs from our \ha\ luminosities using their
formula 17, which is a fifth-order polynomial fit for SFR as a function
of $L_{\rm H\alpha}$.  The coefficients of the polynomial are listed in
their Table 2 for several models, and we used the ones for their
``standard'' model. Their standard model is based on the ``canonical
IMF'' given by their equation 9: a power-law with a
Salpeter slope above 0.5 M\solar and shallower at lower mass. A
comparison between our $SFR_{\rm H\alpha}$ from Table \ref{tab-sfr} and the
Pflamm-Altenburg \et\ $SFR_{\rm PWK}$ is shown in Figure
\ref{fig-comparepwk}. The Pflamm-Altenburg \et\ prescription increases
the \ha-based SFR by a large factor. $SFR_{\rm PWK}$ is higher than our
$SFR_{\rm H\alpha}$ by a factor of 3 at high SFR, increasing to a
factor of 45 at the lowest SFR (because of the very steep galaxy-wide
IMFs used in the PWK model). The $SFR_{\rm PWK}$ are also higher than
our $SFR_{\rm V}$ by a factor of order 3 on average, as shown in
Figure \ref{fig-intsfrpwk}. 
This discrepancy with the observations becomes even worse if we 
compare the PWK rate to $SFR_{\rm FUV}$, which is lower than $SFR_{\rm 
V}$ at low surface brightness (Fig. \ref{fig-intsfr}).
Thus, the PWK prescription gives a star formation rate that is much
higher than either the conventional H$\alpha$ or $V$-band rates, and
these two rates are in better agreement with each other than with the
PWK rates.
In addition PWK predicts a dramatic fall off in the traditional 
$SFR_{\rm H\alpha}$ relative to $SFR_{\rm FUV}$ at low values of the
true SFR (for example, with radius within a galaxy) that we do not see.

Generally we hesitate to interpret the observed variations between
$SFR_{H\alpha}$ and $SFR_{\rm FUV}$ in terms of a variable IMF. The
basic problem is that the IMF also influences many other things, such
as elemental ratios, color evolution, mass-to-light ratios, x-ray
binary and pulsar fractions, and so on, and there is no evidence for
these other variations yet. Another problem is that the correlation in
Figure \ref{fig-imf} is shallow and steady, unlike the correlation
between stellar mass and the ratio of FUV flux to Lyman continuum flux
(shown in Fig. 10 of Meurer et al. 2009), which should produce more of a
threshold effect. 
Finally, the correlation between $SFR_{\rm FUV}/SFR_{\rm
H\alpha}$ and $SFR_{\rm H\alpha}$ for whole galaxies (Figure \ref{fig-imf})
is analogous to
the correlation between $SFR_{\rm FUV}/SFR_{\rm H\alpha}$ and
galactocentric radius in each galaxy (Figure \ref{fig-sfrrat-im}-\ref{fig-sfrrat-nord}),
where at larger radii there is a shift toward lower
$H\alpha$ surface brightness and higher $SFR_{\rm FUV}/SFR_{\rm
H\alpha}$ ratio.

These considerations lead us to believe that the IMF and star 
formation properties in these galaxies are are not abnormal because of 
the PWK cluster-sampling effect. Cluster sampling anomalies also appear 
to be ruled out by the cluster birthline in a plot of 
$L_{\rm H\alpha}/L_{\rm Bol}$ versus $L_{\rm Bol}$ for M33 clusters 
(Corbelli \et\ 2009). Still, there could be other reasons for 
bottom-heavy IMFs in low surface-brightness regions. The pressure and 
density are low in these regions and this may affect the upper stellar 
mass limit or the upper-mass IMF slope in physical ways, 
independently of the star cluster mass or the total number of clusters 
in a galaxy (e.g., Elmegreen 2004). In the PWK model, the upper stellar 
mass depends rigorously on the cluster mass (rather than in an average 
sense only), and the upper cluster mass depends rigorously on the 
number of clusters in the galaxy (again, as distinct from the average 
dependence seen by Whitmore 2003).

In addition, a considerable fraction of the diffuse \ha\ around massive 
stars should be too faint to see when the surface brightness is low 
(Elmegreen \& Hunter 2006).
For example, Melena et al. (2009) found a significant lack of
H$\alpha$ from UV-bright star-forming regions in the outer parts of
a sub-sample of the galaxies studied here. They showed that Str\"omgren radii
around even single massive stars can be comparable to or larger than
the gaseous scale height. Then a high fraction of the Lyman continuum
photons can escape the galaxy and what remains can produce an ionized
medium that is too low in emission measure to observe by present-day
techniques. 
Galaxies with low star formation rates could have a lot of leakage 
because these galaxies have low gas surface densities, low pressures, 
and low densities for the absorption of ionizing photons. They might 
also have more blow out of their interstellar gas near star-forming 
regions because of the low binding energy of the disk. On the other 
hand, if the fraction of star formation in a clustered form is lower in 
low-pressure environments (e.g., Elmegreen 2008; Elias, Alfaro \& 
Cabrera-Cano 2009), then the ISM blow out may be less concentrated in 
dwarfs. 

The correlations between $SFR_{\rm FUV}/SFR_{\rm H\alpha}$ and
$SFR_{\rm H\alpha}$ in Figure \ref{fig-imf} and in Meurer et al. (2009) can
be explained if H$\alpha$ fills the whole disk thickness in the patches
where O-star ionization occurs. We think of this as {\it saturated
ionization}. An important consideration is the gas disk thickness,
which is about constant from galaxy to galaxy. The H$\alpha$ surface
brightness is generally proportional to the emission measure, $n_e^2L$
for electron density $n_e$ and HII region depth $L$. When ionization
fills the disk thickness, $L$ is this thickness and $n_e$ is the
average ISM density, $n_{\rm gas}$. Thus, H$\alpha$ surface brightness
scales with $n_{\rm gas}^2$ if $L$ is about constant. The star
formation rate per unit area determined from H$\alpha$ is directly
proportional to the $H\alpha$ surface brightness, and therefore to
$n_{\rm gas}^2$ too, in this saturated case. Such a $SFR_{\rm
H\alpha}-n_{\rm gas}^2$ relation is not the Kennicutt relation nor is
$SFR_{\rm H\alpha}$ in this case really proportional to the star formation rate. It
is just a measure of the total emission measure in a highly ionized,
and perhaps leaky, patch of the ISM. The star formation rate would be
larger than what is derived from the Kennicutt relation using H$\alpha$
if Lyman continuum photons escape through the top and bottom of the
disk. At the same time, the FUV star formation rate probably scales
with the ISM density to some power, $n_{\rm gas}^\gamma$ for
$\gamma\sim1$ to $\sim1.5$, depending on molecular fraction. This is
the relation discussed by Kennicutt (1998), Wong \& Blitz (2002), Leroy
et al (2009) and others, but written in terms of density rather than
column density; for constant $L$.
Eliminating $n_{\rm gas}$ 
yields $SFR_{\rm FUV}/SFR_{\rm H\alpha}\propto SFR_{\rm H\alpha}^{\gamma/2-1}$. For
$\gamma=1$ or 1.5, $SFR_{\rm FUV}/SFR_{\rm H\alpha}\propto SFR_{\rm
H\alpha}^{-0.5}$ or $SFR_{\rm H\alpha}^{-0.25}$, respectively. These
two power laws span the relationship shown in Figure \ref{fig-imf}.

In the far-outer parts of disks, the thickness should increase in a
typical flare and then the local ISM density should begin to drop more
quickly, making the emission measure around a typical O-type star drop
suddenly as well (Elmegreen \& Hunter 2006) and the escape of Lyman
continuum photons even more likely. 

\subsubsection{Comparison of FUV and $V$}\label{comp}

$SFR_{\rm V}$ is a measure of the average SFR over
a much longer time scale than that measured by FUV.
Figure \ref{fig-intsfr} shows that
$SFR_{\rm FUV}$ and $SFR_{\rm V}$ are comparable on average for most of the galaxies,
but with a large scatter about the relationship.
The equivalence of the SFRs at two different time scales implies that,
on average, the SFR is constant in most dwarf galaxies.
The large scatter around the relationship implies that integrated SFRs may vary 
routinely by factors of a few. This is consistent with the notion of
``gasping'' star formation histories determined from color magnitude diagrams
of nearby dwarfs
in which the star formation rate varies
by factors of a few over long periods of time (for example,
Marconi \et\ 1995). We consider this level of variability in the SFR to
be due to statistical variations in star formation in a small galaxy.

Although $SFR_{\rm V}$ is similar to $SFR_{\rm FUV}$, 
there is one particular uncertainty in $SFR_{\rm V}$ that merits discussion.
The $SFR_{\rm V}$
has been derived by computing the mass in stars and dividing by a
constant age of 12 Gyr. But, Bell \& Bower (2000) and Bell \& de
Jong (2001) find that the best fits to colors and color profiles for
spirals come from models in which the bulk of the disk formation epoch
is a function of the halo mass of the galaxy. 
According to these studies, galaxies with total
masses of order $10^{13}$ M\solar\ have ages of 12 Gyr, but galaxies
with masses of order $10^9$ M\solar\ have effective ages of order 4 Gyr
in the sense that that is  the timescale over which the bulk of stars
have formed. Using an age of 4 Gyr would increase the $SFR_{\rm V}$ by
a factor of 3 or 0.48 in the logarithm of the rate. If generally applied to
all dwarfs, this factor
would then make $SFR_{\rm V}$ larger on average than the current
SFR by a corresponding factor of 3. To explore
this idea, we plot the galactic stellar mass against the ratio of
$SFR_{\rm FUV}$ to $SFR_{\rm V}$ in Figure \ref{fig-ratmv}. The stellar
mass is from $M_V$ and $M/L_V$. We see that the $SFR_{\rm V}$ is too
low relative to $SFR_{\rm FUV}$ by a factor of up to 3 (up to 0.5 in
the log) for some galaxies, and too high by up to a factor of 30 (1.5
in the log) for others. The galaxies requiring higher $SFR_{\rm V}$,
which are mostly systems with masses $\leq 10^9$ M\solar, would benefit
from a lower galactic age. Thus, a lower effective galaxy age for some
systems is possible, although there is no clear trend with mass.
On the other hand, for the galaxies with
$SFR_{\rm V}$ that are too high relative to $SFR_{\rm FUV}$
a much greater age does not solve the problem, and
instead an alternate form for the star formation history would make
more sense, requiring the current SFR to be lower than the average
past rate. We explore the effects of star formation histories on
integrated SFRs in the next section.

\subsection{Color Modeling} \label{sec-modsfr}

We have modeled a simple evolutionary history of our galaxies using
population synthesis fitting of integrated UV, optical, and near-IR
colors, as available, given in Tables \ref{tab-opt} and \ref{tab-phot}.
We used the library of stellar populations from Bruzual \& Charlot
(2003), choosing the ``Padova 1994'' stellar evolutionary tracks
(Alongi \et\ 1993, Bressan \et\ 1993, Fagotto \et\ 1994a,b) and a
Salpeter (1955) stellar IMF. We also used the oxygen abundances of our
galaxies to choose the spectral synthesis models, $Z=0.0004$, 0.004, or
0.008, that are closest to the observed metallicities. If no oxygen
abundance was available, we used the $M_B$ of the galaxy and Richer \&
McCall's (1995) relationship between $M_B$ and $O/H$ for dwarfs to
estimate the proper metallicity choice. The chosen model $Z$ is given
in Table \ref{tab-intmod}.

We fit the colors for each galaxy with a constant SFR model and a
declining SFR with 6 different decay times---$10^{10}$ yrs, $10^9$ yrs,
$7\times10^8$ yrs, $3\times10^8$ yrs, $10^8$ yrs, and $10^7$ yrs.
Declining star formation models assume that the star formation rate was
higher in the past, which is true for the universe in general as a
result of higher relative gas fractions and galaxy interaction rates.
Our modeling program loops over age and finds model colors. For each
model color we subtract the observed color and divide the difference by
the uncertainty in the measured color. We then sum the square of this
difference over the colors. This is a $\chi^2$. Among all possible
solutions, we then average with an exp($-0.5\chi^2$) weighting factor
to obtain a final age and mass. The uncertainties in many of our colors
are sufficiently small compared to the model SED fits that the $\chi^2$ values were unreasonably
large (often greater than 40). So, we multiplied all of the color
uncertainties by a constant factor of 20 for all galaxies in order to get
non-zero Gaussian weights. This is effectively the same as considering
greater than statistical uncertainties, or systematic uncertainties, in
the observations and models. In a small number of cases the $\chi^2$
method does not find a good solution at all. We considered this to be
the case when 5 or fewer of the age models out of the 164 trial ages in
the Bruzual \& Charlot tables gave
$\chi^2<40$. We then modified our method to obtain the best possible
SED fit by averaging together the trial ages and masses from only those
age trials that had the lowest rms deviation between the observed and
modeled colors (i.e., not dividing by the measurement errors as in the
$\chi^2$ method). The weights for this average were taken to be
Gaussian functions of this rms deviation, rather than $\chi^2$. The
uncertainties in the results for these cases were taken from the rms
values in the averages.  Of the 7 models for each galaxy,
we chose the one that best fit the galaxy colors.

The star formation
rate SFR$_{\rm Col}$ is taken to be the color-derived mass in stars
divided by the color-derived age of the galaxy for the best-fit model.
We correct the model mass for the mass recycled back into the interstellar medium
by increasing the mass by a factor of two
(Brinchmann \et\ 2004). 
The best fit model parameters are given in Table \ref{tab-intmod}.

The color-derived age is approximately the time when the most prominent
burst of star formation occurred. The corresponding SFR is the average
since that time. Generally we expect $SFR_{\rm Col}$ to be larger than
$SFR_{\rm V}$ because any burst reflected by $SFR_{\rm Col}$ will be
stronger than the average rate over a Hubble time, as reflected by
$SFR_{\rm V}$. Similarly, $SFR_{\rm Col}$ will be larger than $SFR_{\rm
FUV}$ if there has been some decay in the SFR over the last several
Gyr.

A histogram of the color-derived ages is shown in Figure
\ref{fig-histmodages}. The ages cluster around 1 Gyr, with a few
larger values for Im and Sm types than for BCD types. This Gyr
color-derived age means that most of the optical light in these
galaxies comes from star formation within the last Gyr. 
The fact that BCDs have color ages that are similar to those of the 
other types implies that BCDs are not undergoing a recent and global 
burst, lasting only 100 Myr for example. The intense inner disk 
activity in BCDs may be somewhat recent, as suggested by its relatively 
blue color, but the whole disk SFR has to be sustained for a Gyr or 
more. This means that the Im and Sm types differ from BCDs primarily in 
the former's lack of spatial concentration for the most recent star 
formation.

In Figure \ref{fig-sfrintcol} we compare SFR$_{\rm Col}$ to the SFRs
determined from the $V$-band and FUV. We see that SFR$_{\rm Col}$ is
indeed greater than both of these SFR measures -- by a factor of about
5 on average. As discussed above, this factor follows from the
exponential star formation history that we have fit to the optical
colors. With this history, the instantaneous star formation rate is
$S(t)=S_0\exp\left(-t/t_d\right)$ for rate normalization $S_0$, time
$t$ from some past beginning, and decay time $t_d$. The total mass is
$\int_0^{\rm age}S(t)dt$, and the average rate is this mass divided by
the age. Writing $\xi={\rm age}/t_d$, the average rate becomes
$S_0\left(1-e^{-\xi}\right)/\xi$. The rate today is much smaller than
this, $S_0e^{-\xi}$. The ratio of the past average divided by the
today's rate is $\left(e^\xi-1\right)/\xi$. Figure \ref{fig-sfrintcol}
indicates a value of $\sim5$ for this SFR ratio, which implies
$\xi\sim2.7$ or $\log \xi\sim0.43$. The average value of $\log \xi$ in
Table \ref{tab-intmod} is 0.67$\pm$0.25.  The similarity between this
average observed value of $\log \left({\rm age}/t_d\right)=0.67$ and
the value required to give the observed $SFR_{\rm Col}/SFR_{\rm
FUV}\sim5$ is to be expected from the model.

Different star formation histories will produce slightly different
offsets between the history-derived SFRs and the $V$-band or FUV-band
SFRs. Star formation in small galaxies is likely to be bursty and
gaspy, not steady or purely exponential as the models assume. Still,
our main results should be robust: most dwarf irregulars had a period
of more intense star formation in their 1-Gyr past than they do today,
and this burst also exceeded the past average rate over the history of
the galaxy.  More finely tuned models should be able to find sub-bursts
within this Gyr period. In our recent study of discrete star-forming
regions in these same galaxies, we found factor-of-two variations in
the SFRs on timescales ranging from 10's of millions of years to 1 Gyr
using the age and mass distributions of the regions. The purpose of the
present study is not to reproduce these bursty substructures in any
detail using galaxy-wide colors, but only to illustrate the dominance
of $\sim$Gyr-old star formation to the overall appearance of these
galaxies. This Gyr dominance is not obvious at first when faced with
the FUV images of 100-Myr old star formation and H$\alpha$ images of
10-Myr old star formation. The colors in fact suggest that these more
recent events are relatively small compared to other events that
typically occurred in the last Gyr.

\section{Radial Variations from Azimuthally-averaged Star Formation Rates} \label{sec-radsfr}

From the azimuthally-averaged surface photometry, we calculate SFRs
from the FUV, \ha, and $V$-band luminosities using the same
prescriptions described in section \S 5. However, here the SFRs are in
units of M\solar yr$^{-1}$ kpc$^{-2}$ and correspond to elliptical
annuli of equal width and step size in increasing distance from the
center of the galaxy. The ellipse parameters are given by Hunter \&
Elmegreen (2006). For a fourth SFR measure, we also modeled the colors
of the annuli using the process described for the integrated
photometry.  
For the annuli models we assumed the best model type determined from 
the integrated SFR model, that is, constant star formation or decaying SFR. 
The model type is listed in Table 7 by decay timescale. The model fits 
to the observations were generally better for the annuli than for the 
integrated galaxy, so the $\chi^2$ values were much lower. We therefore 
multiplied all of the color uncertainties by a factor of 5 instead of 
20 in order to get non-zero Gaussian weights.

Radial profiles of all three of our luminosity-based SFR measures and
the model SFR are plotted in Figures
\ref{fig-sfrrat-im}--\ref{fig-sfrrat-nord} as the ratio of $SFR_{\rm
FUV}$ to $SFR_{H\alpha}$ (red curves), to $SFR_{\rm V}$ (blue curves)
and to $SFR_{\rm Col}$ (green curves). The ratios for Im types in
Figure \ref{fig-sfrrat-im} tend to be flat or falling with radius. The
ratios for BCDs at the top of Figure \ref{fig-sfrrat-bcdsm} are mostly
falling, and for Sm types at the bottom of Figure
\ref{fig-sfrrat-bcdsm} are mostly flat. In addition, $SFR_{\rm Col}$
is always higher at each radius relative to $SFR_{\rm FUV}$ than are
the other two SFR measures, consistent with the findings of section
\ref{sec-modsfr}.

The rising or falling trends in Figures
\ref{fig-sfrrat-im}--\ref{fig-sfrrat-nord} are quantified in the
histogram given in Figure \ref{fig-histrat}. Radial gradients in the logarithm
of $SFR_{\rm FUV}/SFR_{\rm H\alpha}$ and the logarithm of $SFR_{\rm
FUV}/SFR_{\rm V}$ were fit with a straight line, and in this figure we
plot the number distribution of galaxies as a function of the slope of
these gradients, normalized to the $V$-band disk scale length $R_D^V$.
Some galaxy $SFR_{\rm FUV}/SFR_{\rm V}$ profiles were fit with two parts
(9 dIm, 4 Sm),
and in those cases the outer gradient is counted here.
A flat radial distribution has a gradient of zero and is marked with a
dashed vertical line in Figure \ref{fig-histrat}. A positive gradient
means that $SFR_{\rm FUV}$ becomes more dominant with radius than
$SFR_{\rm H\alpha}$ or $SFR_{\rm V}$, and a negative gradient means
that $SFR_{\rm FUV}$ becomes less dominant with radius. Figure
\ref{fig-histrat} indicates that most of the galaxies have little or no 
gradientin $SFR_{\rm FUV}/SFR_{\rm H\alpha}$, meaning that the UV and 
\ha-based SFRs on 
average track each other where both are measurable 
(i.e., where the \ha\ and FUV coexist). However, 
$SFR_{\rm FUV}/SFR_{\rm V}$ gradients tend to be negative 
for the dIm and BCD types, while they are closer to zero for the Sm 
types.

The trends in Figures \ref{fig-sfrrat-im}--\ref{fig-histrat} follow
from the other trends shown in this paper, and from the definitions of
the various star formation rate indicators. BCDs have starbursts in
their centers, so the $SFR_{\rm FUV}$ to $SFR_{\rm V}$ ratio tends to
decrease with radius. The same is true for many Im galaxies, but not
for the Sm's, which can have FUV spots relatively far out in their
disks. 
The generally low values of $SFR_{\rm FUV}/SFR_{\rm V}$
most likely follow from higher past star formation rates.

The similarity in radial gradients for $SFR_{\rm FUV}$ and 
$SFR_{\rm H\alpha}$ shown on the 
right in Figure 17 suggests that the IMF is not getting steeper with 
radius over the radial range where both FUV and \ha\ are observable. 
This is in agreement with the conclusions of section 5.1.1.2 where we 
suggested that the missing \ha\ in the outer regions is from faintness. 
It would be unlikely to see no difference in the two SFR gradients 
until there is a sudden lack of \ha\ if the IMF were smoothly 
varying with surface brightness.

To see if there is any connection between the UV color gradient and the
gradient in SFR, we plot Figure \ref{fig-gradgrad}. 
The normalized $\log SFR_{\rm FUV}/SFR_{\rm V}$ 
gradient is plotted in
Figure \ref{fig-gradgrad} along the x-axis, and the normalized FUV$-$NUV color gradient 
along the y-axis.
High values on the y-axis indicate that the disk gets
redder with increasing radius. 
The color profile, like the SFR ratio gradient, was sometimes fit with separate
components for the inner and outer disk (5 dIm, 1 BCD).
For galaxies for which both quantities were fit with two components, the
inner pair are plotted and the outer pair of values are plotted.
For the cases where one quantity was fit with two components and the other was fit with one,
both components of the two-part fit were plotted against the inner component of the 
single-part fit.
Inner profile fits are denoted by filled symbols and outer profiles are denoted by 
open symbols. 
The plotted points show a wide range of
values. There are many galaxies with no color gradient but with a wide
range in SFR ratio gradients. There are also galaxies with small
gradients in the SFR ratio but prominent color gradients. Galaxies with
gradients in both the SFR ratio and UV color tend to show a wide range
in color gradients for each SFR ratio gradient. Still, there is a trend
in that galaxies with both gradients tend to get redder with radius and
also have decreasing $SFR_{\rm FUV}$ relative to $SFR_{\rm V}$ with
increasing radius.  These trends suggest that {\it sometimes} star formation
stops in the outer disk for a relatively long time.

One possible reason for outer disk truncation in star formation is a
lack of gas. We do not yet have HI maps of these galaxies, so we cannot
make a definitive statement on this issue. However, we do have total HI
masses, and these are plotted in Figure \ref{fig-normalizemhilb2}. The
abscissa has the log of the ratio of total HI mass to $B$-band
luminosity, and the ordinate has the normalized SFR-ratio gradient in
the bottom panel and the normalized UV color gradient in the top panel.
In this case, the SFR-ratio gradient is that only for the outer disk,
if fit with two parts.
Evidently, low $M_{\rm HI}/L_{\rm B}$ does correspond to
both an outer disk reddening and an outer disk drop in $SFR_{\rm FUV}$ relative
to disk mass. These cases presumably have halted outer disk star
formation because of a lack of outer disk gas. Above $\log M_{\rm
HI}/L_{\rm B}\sim-0.5$, however, this correlation disappears. Galaxies
with relatively high HI mass can have either large or small SFR-ratio
gradients, although most of them have small UV color gradients.  Such
variations in SFR-ratio gradient should correlate with color gradients
if the SFR gradients in the disk are long term.  Recall that $SFR_{\rm
H\alpha}$ measures star formation in the last few tens of Myr, $FUV-NUV$
measures star formation in the last hundred Myrs or so, and $SFR_{\rm
V}$ measures the average SFR over a Hubble time. Thus, a more likely
explanation for the behavior of $SFR_{\rm FUV}/SFR_{\rm V}$ and
$FUV-NUV$ at high $M_{\rm HI}/L_{\rm B}$ is a short time variation in
the outer disk star formation rate. Observations that are sensitive to
the time scale for variations, such as $SFR_{\rm FUV}$, will mimic
those variations and show a lot of scatter, while observations that are
sensitive only to longer-term variations, such as $FUV-NUV$ or
$SFR_{\rm V}$, will average over the short term effects and show only a
constant or slowly varying rate. Thus, we suggest that galaxies with low
$M_{\rm HI}/L_{\rm B}$ have essentially stopped their star formation in
the outer disk, 
while galaxies with intermediate and high $M_{\rm
HI}/L_{\rm B}$ have variable star formation in the outer disk with a
time scale of several tens of millions to a hundred million years.

In Figure \ref{fig-gradmv} we plot the slope of the normalized gradient
in $\log SFR_{\rm FUV}/SFR_{\rm V}$ against galactic $M_V$. 
There is a rough correlation in the sense that galaxies with more negative
gradients are fainter.  We do not find an equally clear correlation
between $FUV-NUV$ color gradient and $M_V$ (although there is the usual
correlation between $M_{\rm HI}/L_{\rm B}$ and $M_V$ in the sense that
smaller galaxies have relatively more HI). Figure \ref{fig-gradmv} is
consistent with a model in which small galaxies have more time
fluctuations in the star formation rate.

We note that observations that are sensitive to a certain time interval
for star formation will show fluctuations only over the radial range in
a galaxy where the variability has a comparable timescale. Generally,
the density of interstellar gas decreases with radius and the timescale
for SFR variations increases. 
Thus, observations like FUV that are sensitive to $\sim100$ Myr old 
stellar populations will show the strongest fluctuations in the outer 
disk, where the dynamical timescale is larger than this.

\section{Discussion} \label{sec-discuss}

\subsection{Results}

\begin{figure}
\epsscale{1.0}
\plotone{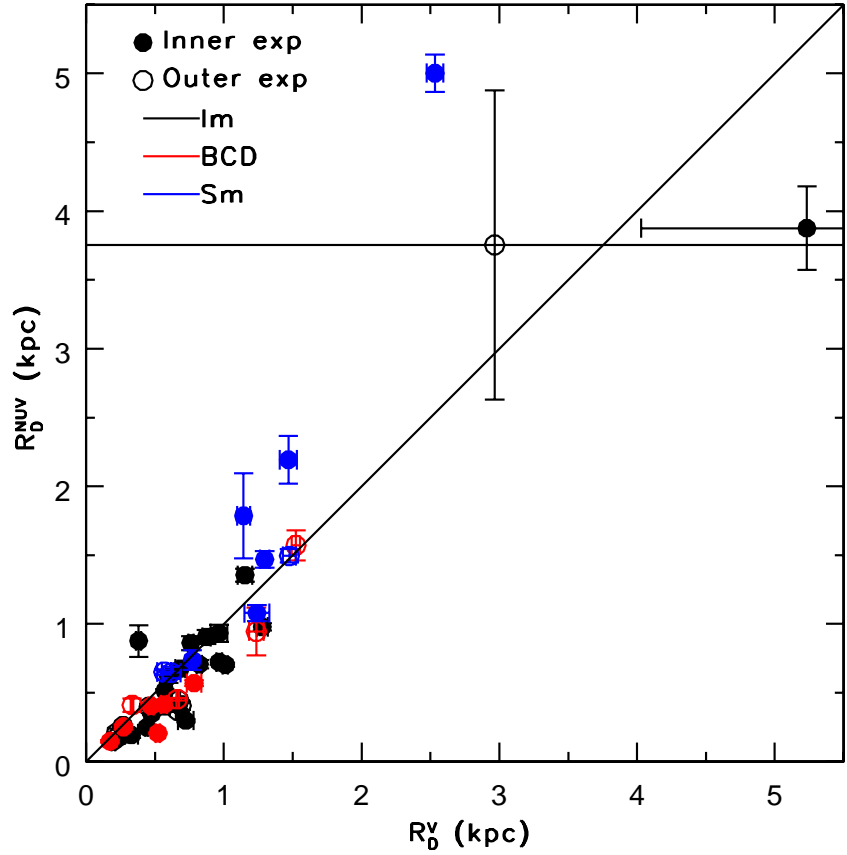}
\caption{Disk scale length determined from NUV surface photometry compared
to those determined from $V$-band surface photometry. For those profiles that were fit
in two parts, the inner exponential is shown as a {\it solid} symbol and the
outer exponential disk scale length is shown as an {\it open} symbol.
The Im sample is shown in {\it black}, BCDs as {\it red}, and Sm galaxies as {\it blue}.
\label{fig-rd}}
\end{figure}

\begin{figure}
\epsscale{1.0} \plotone{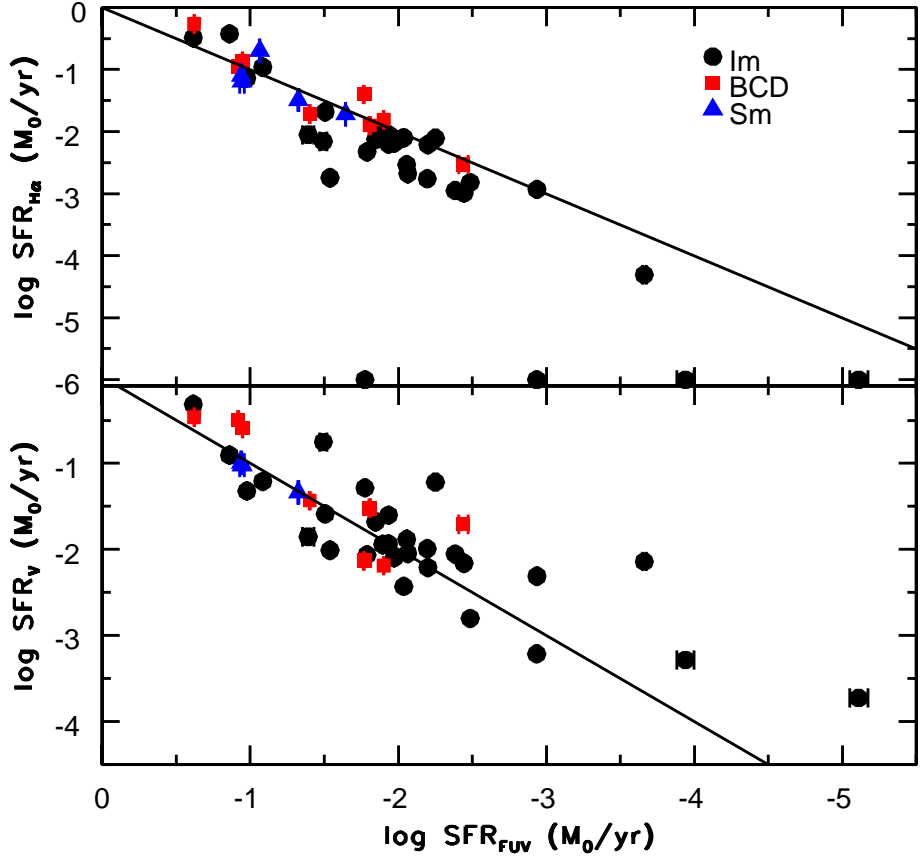} 
\caption{Integrated
star formation rate (SFR) determined from the FUV luminosity
$SFR_{\rm FUV}$ plotted against the SFR determined from the $V$-band
luminosity $SFR_{\rm V}$ and from the \protect\ha\ luminosity $SFR_{\rm
H\alpha}$. $M_V$ is converted to mass in stars using a stellar
$M/L_V$ ratio, which depends on $(B-V)_0$, and the $SFR_{\rm V}$
assumes a constant SFR over 12 Gyr. Galaxies with no \protect\ha\
emission are plotted at $\log SFR_{\rm H\alpha}=-6$. The solid black
lines denote equal SFRs. The $SFR_{\rm FUV}$ are higher than $SFR_{\rm
H\alpha}$ for lower SFR systems, and lower than $SFR_{\rm V}$ for
lower SFR systems. \label{fig-intsfr}}
\end{figure}

\begin{figure}
\epsscale{1.0} \plotone{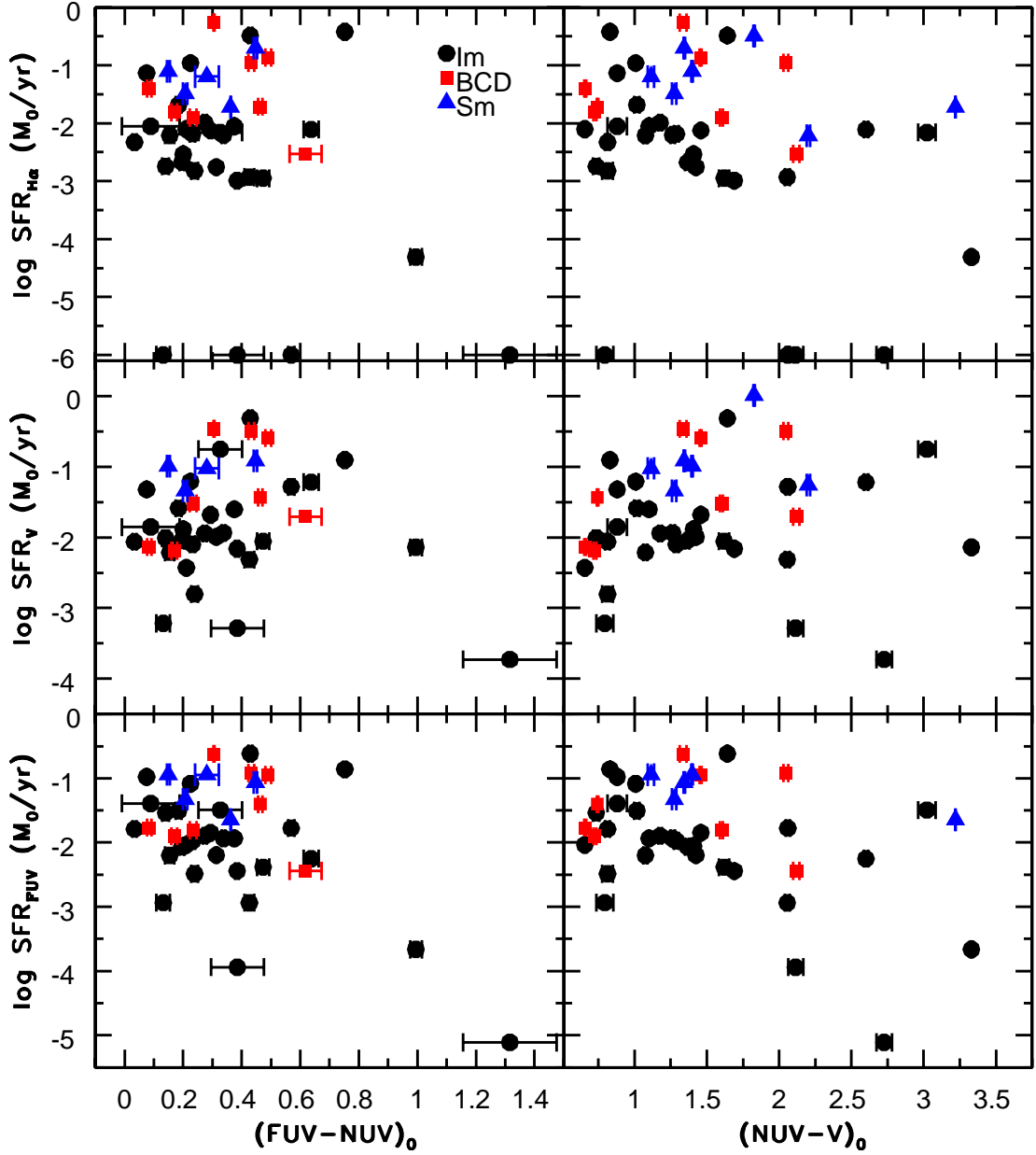} 
\caption{Integrated
SFRs determined from \protect\ha, the $V$-band luminosity, and the
FUV luminosity plotted against the UV color ({\it left}) and
UV-optical color ({\it right}). Galaxies with no \protect\ha\ emission
are plotted at $\log SFR_{\rm H\alpha}=-6$ ({\it top}). There is no
trend of SFRs with these integrated galactic colors.
\label{fig-sfrcolor}}
\end{figure}

\begin{figure}
\epsscale{1.0}
\plotone{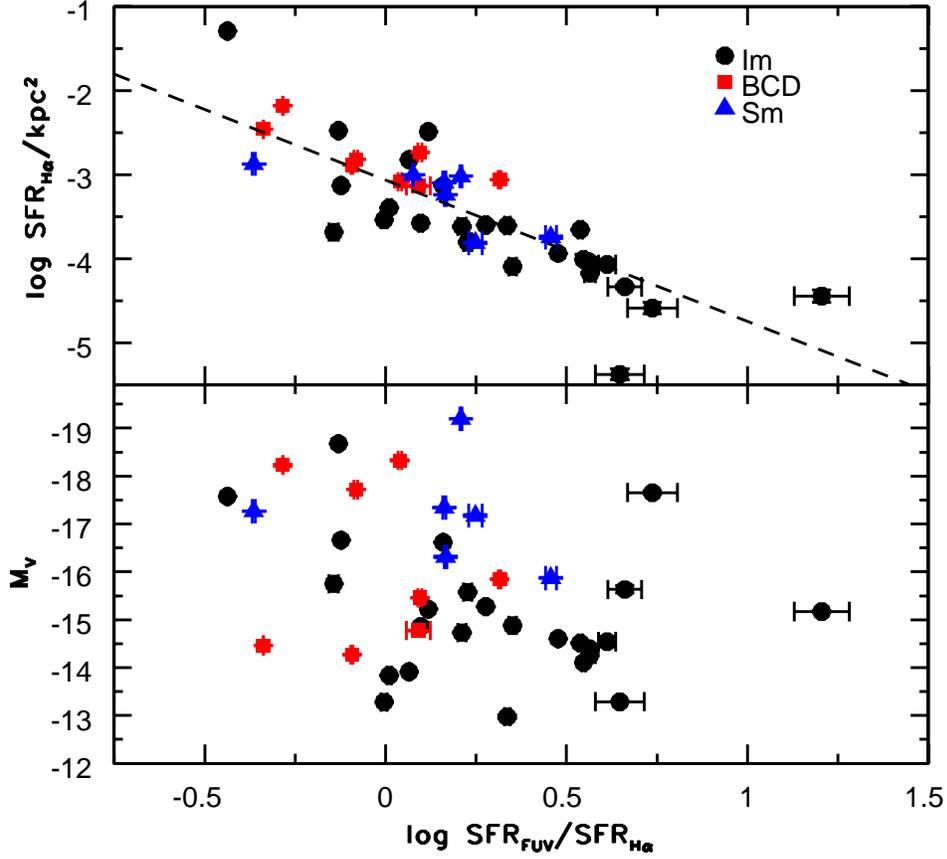}
\caption{Ratio of integrated SFRs measured from the FUV to SFRs measured from
H$\alpha$ vs. the H$\alpha$ SFR per unit area ({\it Top})
and galactic $M_V$ ({\it Bottom}).
The area used to normalize $SFR_{\rm H\alpha}$ is that over which H$\alpha$ was measured.
The dashed line in the upper panel is a fit to all but the three outlying points:
$\log SFR_{\rm H\alpha}/{\rm kpc^2} = -3.06\pm0.07 - (1.68\pm0.20)\times\log SFR_{\rm FUV}/SFR_{\rm H\alpha}$.
\label{fig-imf}}
\end{figure}

\begin{figure}
\epsscale{1.0}
\plotone{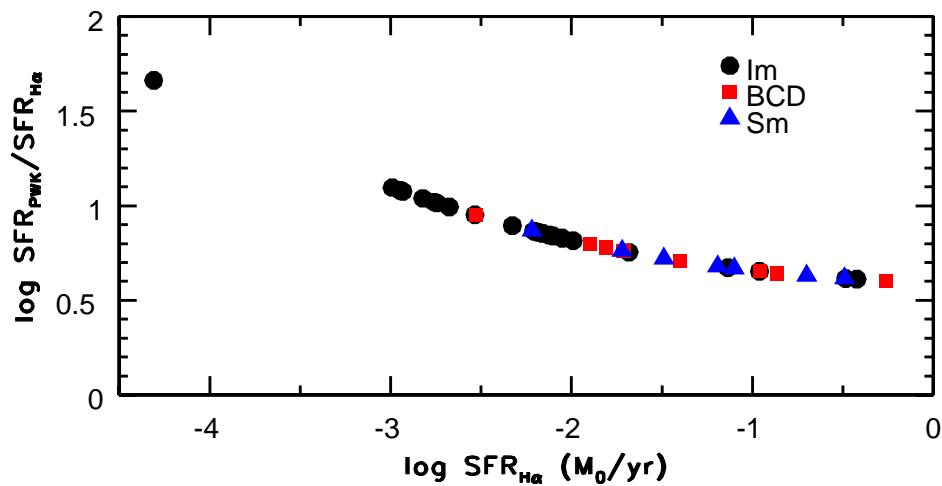}
\caption{Ratio between integrated star formation rates determined from \protect\ha\
luminosities $SFR_{\rm H\alpha}$, as given in Table \protect\ref{tab-sfr}, and from the prescription
of Pflamm-Altenburg, Weidner, \& Kroupa (2007) $SFR_{\rm PWK}$. The Pflamm-Altenburg \protect\et\
formula is a function of the SFR itself, and the difference between our
adopted conversion method and that of Pflamm-Altenburg \protect\et\ increases as
the SFR decreases, with $SFR_{\rm PWK}$ being higher than $SFR_{\rm H\alpha}$ by factors of
up to 45.
\label{fig-comparepwk}}
\end{figure}

\begin{figure}
\epsscale{1.0}
\plotone{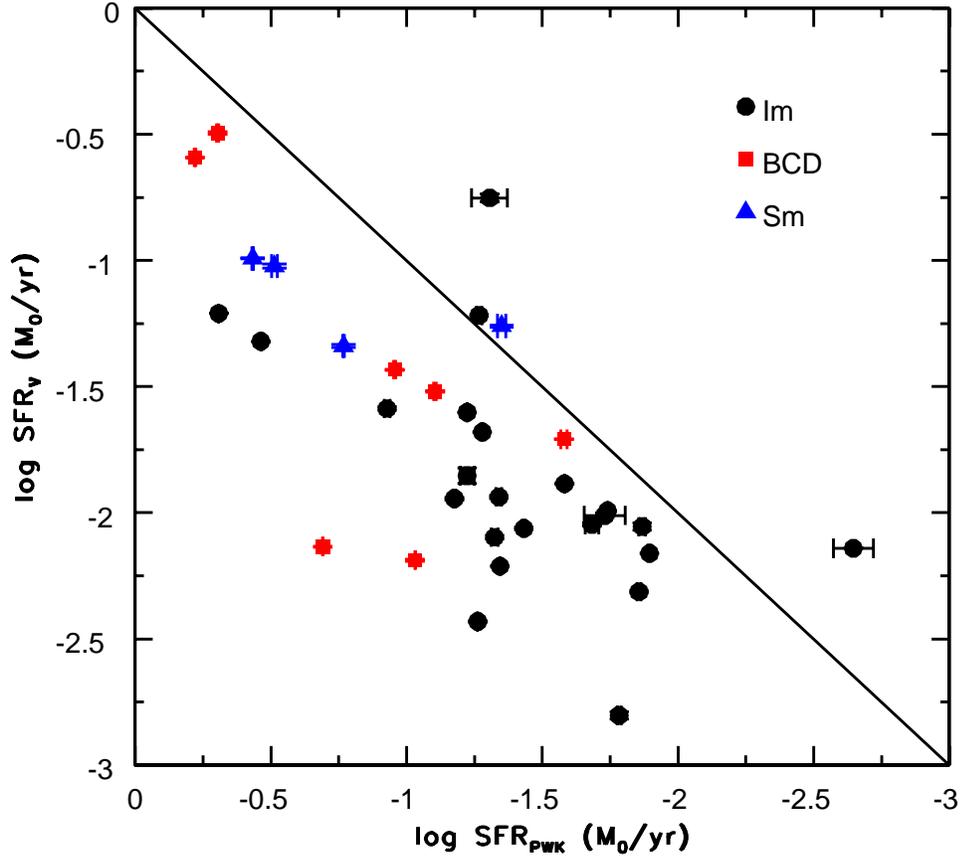}
\caption{Comparison of the star formation rate determined from \protect\ha\ luminosities
following the prescription of Pflamm-Altenburg, Weidner, \& Kroupa (2007) $SFR_{\rm PWK}$
and the SFR determined from $M_V$ $SFR_{\rm V}$. The slanting solid
line marks equal SFRs.
The Pflamm-Altenburg et al.\ prescription exacerbates the discrepancy between between
the \protect\ha-based SFR and the SFR determined from $M_V$.
\label{fig-intsfrpwk}}
\end{figure}

\begin{figure}
\epsscale{1.0}
\plotone{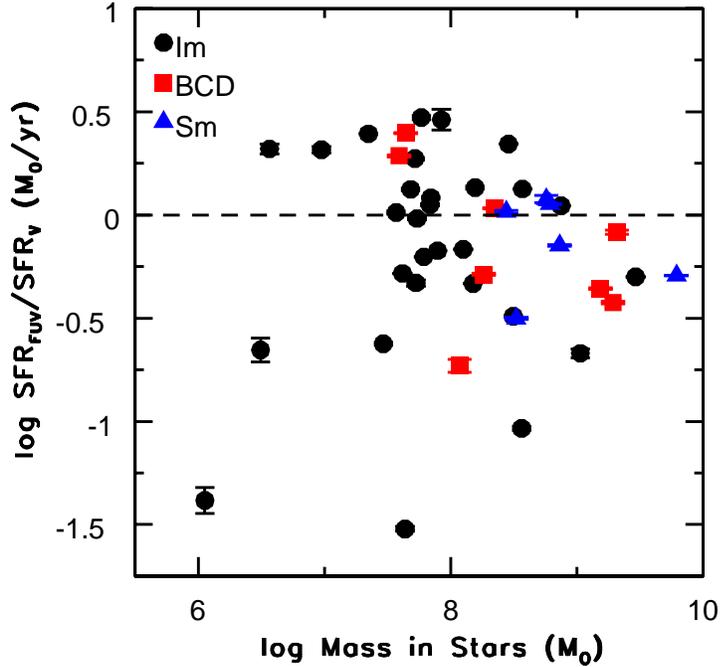}
\caption{Ratios $SFR_{\rm FUV}/SFR_{\rm V}$ as a function of integrated galactic
mass in stars. The mass in stars is determined from $M_V$ and $M/L_V$ as described
in the text.
The dashed horizontal line marks a ratio of 1.
This plot is used to explore the possibility that the SFRs based on total mass over estimate the bulk
age of the galaxy as a function of mass of the galaxy, an idea proposed by
Bell \& Bower (2000).
\label{fig-ratmv}}
\end{figure}

\begin{figure}
\epsscale{1.0}
\plotone{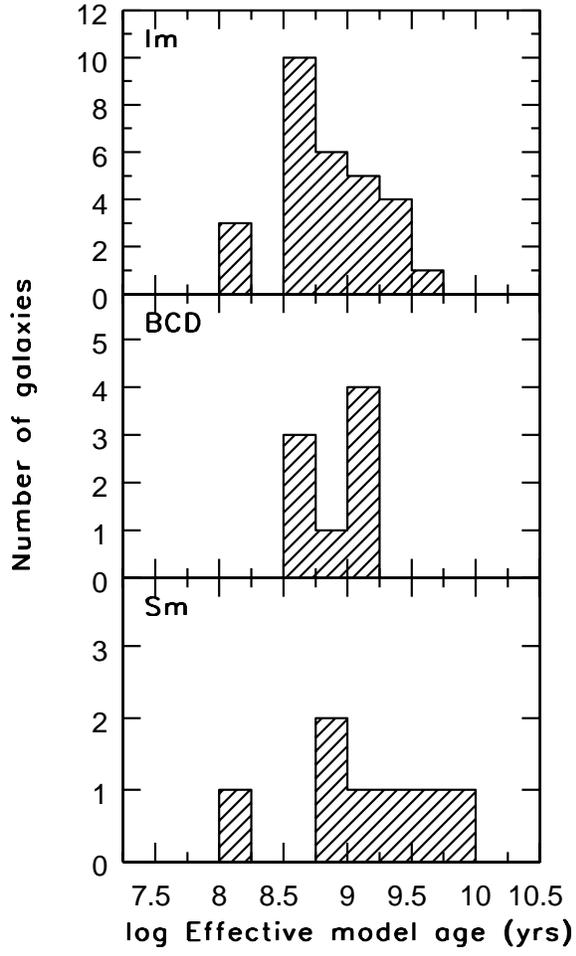}
\caption{Histogram of effective galactic ages from models of the integrated colors.
\label{fig-histmodages}}
\end{figure}

\begin{figure}
\epsscale{1.0}
\plotone{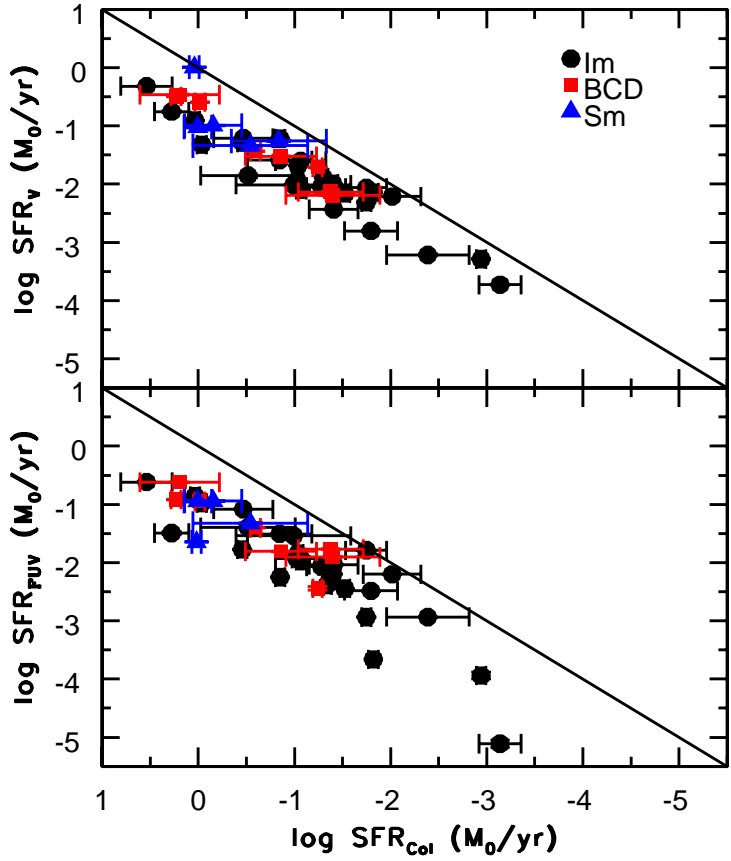}
\caption{SFR determined from population synthesis of integrated UV, optical, and near-IR
colors $SFR_{\rm Col}$ plotted against SFRs determined from $V$-band and FUV luminosities.
The solid line denotes equal SFRs. $SFR_{\rm Col}$ is greater than the other two SFR measures.
\label{fig-sfrintcol}}
\end{figure}

Observations at FUV, NUV, $V$, and H$\alpha$ of integrated magnitudes,
radial profiles, and star formation rates for dIm, BCD, and Sm
galaxies reveal star formation in outer disks that is a smooth
continuation of the inner disk activity, with a bursty nature on a
timescale of $\sim1$ Gyr.  Surface brightness levels are observed down
to $\sim29$ mag arcsec $^{-2}$ in NUV and $\sim27$ mag arcsec $^{-2}$
in $V$ (Fig. \ref{fig-profiles}).  The $(FUV-NUV)_0$ and $(NUV-V)_0$ colors are
usually red in the outer disk, and often increasingly red with
galactocentric radius, indicating a slowdown of star formation in the
outer parts over the last Gyr.  The BCD and Im galaxies, which are
physically small ($R_D^V\sim0.2-1$ kpc), tend to be the ones with red
outer disks, and they have $V$-band scale lengths larger than 
those measured in the NUV-band by
a factor of $\sim1.2$ (Fig. \ref{fig-rd}). The Sm galaxies, which can
be physically large ($R_D^V\sim0.5-3$ kpc), tend to have blue outer
disks and $V$-band scalelengths that are smaller than 
those measured in the NUV-band.

These color trends are reproduced by trends in star formation
indicators. Short-term indicators like H$\alpha$ behave differently
with radius than intermediate and long-term indicators like FUV and
$V$-band, respectively.  
Galaxies that get redder with radius tend to have
radially decreasing FUV relative to $V$ star formation rates, while
galaxies that get bluer with radius tend to have radially increasing
FUV relative to $V$ star formation. 
The smallest integrated star formation rates are on the order of
$10^{-3}-10^{-4}\;M_\odot$ yr$^{-1}$. The smallest areal star formation
rates are on the order of $10^{-4}-10^{-5}\;M_\odot$ yr$^{-1}$
kpc$^{-2}$ (Fig. \ref{fig-imf}), as measured by H$\alpha$. At these
rates, the surface density of stars formed after a Hubble time is only
$\sim1\;M_\odot$ pc$^{-2}$ on average inside a $\sim4$ kpc$^{2}$
disk, the typical size for our galaxies.  The stellar surface densities
are much lower in the outer parts. We return to this point below.

Double exponential disks, with steeper or shallower outer parts than
inner parts, are observed in this sample (Fig. \ref{fig-profiles}).
Usually the NUV follows the $V$-band, where double exponentials were
already observed (Hunter \& Elmegreen 2006).  There is often no obvious
$(FUV-NUV)_0$ color change at the break radius (e.g., DDO 68, DDO 75, F565V2,
DDO 150), although sometimes there is (DDO 101, DDI 183, M81DwA), and
sometimes there is an abrupt color change without a break in the radial
profile (e.g. IC 1613).  This ambiguity in color breaks, in addition to
the lack of spiral density waves in our sample, suggests that the outer
exponential is not the result of spiral wave-scattered stars, as
proposed for spiral galaxies by Roskat et al. (2008). No disk
terminations have been observed in our sample, even down to the
faintest levels and largest radii obtained, which is usually with the
$V$ and NUV observations.

Generally, the outer disks of galaxies typically lack H$\alpha$ emission even
though there is star formation at some level seen in the FUV. The same is
true for our sample. We find a smooth correlation between the ratio of
FUV to H$\alpha$ star formation rates versus the absolute star
formation rate, of the form $SFR_{\rm FUV}/SFR_{\rm H\alpha}\propto
SFR_{\rm H\alpha}^{-0.59\pm0.07}$. This is essentially the same
correlation as that found by Meurer et al. (2009) for different
galaxies.   
We disagree with the
conclusions in Meurer, et al., however. They considered this
correlation to be the result of a steepening stellar IMF with
galactocentric radius, so that outer disk star formation lacks O-type
stars and the associated ionization compared to inner disk star
formation. 
However, pervasive FUV in the outer disks and the radially
invariant $FUV-V$ colors in many cases suggest that the IMF does not
change with radius and the missing H$\alpha$ is from a lack of gas,
i.e., from density-bounded HII regions in the classical sense. 
At the very least, this says that the stars that produce the $FUV-V$ colors
are not varying, but the range of masses responsible for \ha\ emission could be.
At very
low ambient densities, this HII region limitation, which is usually in
reference to local conditions near giant molecular clouds, translates
into a limit for the whole thickness of the local disk (``saturated
ionization''). We showed in \S \ref{fuv-ha} that for constant disk
thickness over the radial range where H$\alpha$ is observed, and for a
local star formation rate proportional to $n_{\rm gas}^\gamma$, the
observed ratio becomes $SFR_{\rm FUV}/SFR_{\rm H\alpha}\propto SFR_{\rm
H\alpha}^{\gamma/2-1}$.  Typically $\gamma\sim1$ to 1.5, and the
observed slope of this correlation is reproduced.  Beyond the radial
range where H$\alpha$ is observed, the disk thickness probably flares,
making the ionization even more diffuse and any associated H$\alpha$
even harder to observe. Thus, we believe H$\alpha$ is lacking because
the gas density is too low to show the associated emission measure. We
predict that more sensitive observations of H$\alpha$ or other
ionization tracers over large regions around each star formation site
will show the presence of ionization.

Models for the color profiles of our galaxies considered an
exponentially decaying star formation history with a constant IMF. The
resulting average star formation rates were compared to the $H\alpha$,
FUV and $V$-band rates. There is a general agreement between all of these
rates if the star formation rates were higher in the past and the
star-formation ages or burst ages in the outer disks are $\sim1$ Gyr or a
few Gyr.  A higher rate in the past also helps to explain how the FUV
profiles can be steeper than the $V$-band profiles, especially in Im
galaxies (Fig. \ref{fig-profiles}). Current star formation following
steep FUV profiles cannot continue building a disk with the same mass
distribution indicated by the shallow $V$-band profiles. The disk scale
lengths will get smaller over time if the FUV profiles alone reflect a
long future of star formation.  Outer disk bursts with $\sim$Gyr
timescales mitigate this problem, allowing the disks to build up in a
self-similar although chaotic way, or even with increasing disk scale
lengths over time. The Sm galaxies with relatively blue outer disks and
high outer disk star formation rates could be examples of outer disk
building.  Perhaps the morphology changes from Im to Sm during this
outer active phase, because the Sm types have more extended and bluer
outer disks compared to the Im types.

There is a correlation between the outward red $FUV-V$ gradients and the
HI mass-to-light ratio of the galaxy in the sense that galaxies with
the lowest relative HI mass have the largest scale-normalized gradients
in the ratios of FUV to $V$-band star formation rates (Fig.
\ref{fig-normalizemhilb2}). We interpret this to imply that some outer
disks are red because star formation virtually stops there with a lack
of gas. Galaxies with intermediate to high relative HI masses have a
wide range of gradients in the star formation rate ratio, which implies
that outer disk star formation is ongoing, but bursty, as indicated by
our other results.

Perhaps the most remarkable result of this survey is the low level of
FUV emission in the far-outer disks of late-type and dwarf irregular
galaxies.  Star formation is apparently continuing there at an
extremely low level.
In the next subsection, we model the population and surface
density in the far outer parts of the disks, and consider the
implications of these models for star formation and galaxy buildup.

\subsection{Models for Far Outer Disk Colors and Surface Brightnesses}

Figure \ref{fig-profiles} summarizes most of the key observations in
our study and indicates several main features of outer disks. First the
$FUV-V$ colors are usually in the range of $\sim1$ to $\sim3$ mag, 
which are typical of star-forming galaxies (Wyder \et\ 2007).
Nevertheless, the surface brightness level in the
FUV also gets very low, down to $\sim29$ mag arcsec$^{-2}$ in some
cases, with no indication of a turnover. In $V$-band, the surface
brightness may reach $\sim26$ to $\sim28$ mag before it gets too faint
to see. These colors and surface brightnesses place interesting
constraints on the stellar populations and surface densities in the
outer disk.

We consider again the models of Bruzual \& Charlot (2003) for the
Salpeter IMF at two metallicities, 0.4 solar and 0.2 solar. 
Figure
\ref{fig-mag22} shows the absolute magnitudes at 2267\AA, the effective
wavelength of the NUV filter, and $V$ for a
single stellar population (SSP) model starting with $1\;M_\odot$ of
stars (red and blue curves). 
The populations get fainter as they age,
with the FUV getting faint faster than the $V$-band; the cross over
occurs at an age of $\sim10^{7.4}$ years.  
The black curves are the $2267-V$ colors for a constant star formation 
rate; this model also gets redder with age, but much slower than the 
SSP model: only after $10^{8.2}$ yrs does the $2267-V$ color become 
positive and it never gets much above 1, unlike the observations in 
Figure \ref{fig-profiles}.
Thus, the observation
of red colors in the outer disks of our sample, along with the
observation of steeper FUV profiles than $V$ profiles, supports the SSP
model more than the constant SFR model. That is, star formation is
bursty, particularly in the outer disks, with an off time comparable to
or larger than 1 Gyr, which is what it takes to make $2267-V$ positive.

Figure \ref{fig-mag22-sigma} shows the surface density of stars versus
age in the SSP (blue, red, magenta curves) and constant SFR models
(black curves) for a surface brightness in the FUV of $\sim29$ mag
arcsec$^{-2}$ (blue curves) and two sample surface brightnesses in the
$V$-band, $\sim28$ mag arcsec$^{-2}$ (red) and $\sim29$ mag arcsec$^{-2}$
(magenta).  A green square outlines the intersection point where FUV
and $V$ surface brightnesses give the same mass surface density, as
required for a sensible model. 
The models with a constant star formation rate are generally too blue 
to produce a sensible model: the $M_{2267}=29$ mag arcsec$^{-2}$ black curve never 
intersects the $M_{\rm V}=27$ mag arcsec$^{-2}$ black curve and it only intersects the 
$M_{\rm V}=28$ mag arcsec$^{-2}$ black curve at a very late time, $10^{9.4}$ yrs.
The results suggest that the age of the stellar population in a 
typical outer disk is several hundred Myr, without much star formation 
in the intervening time. 
This is consistent with the ages obtained from other
observations and models in this paper. Also, the surface density in the
outer disk is $\sim0.1\;M_\odot$ pc$^{-2}$.  This is the lowest surface
density that we measure here, corresponding to the faintest FUV surface
brightness. Presumably surface density goes even lower at larger radii.

A surface density of $\sim0.1\;M_\odot$ pc$^{-2}$ corresponds to a gas
column density of $\sim10^{19}$ cm$^{-2}$, which also corresponds to
0.01 mag of extinction at Solar abundances. This is extremely low
compared to the main disks of spiral galaxies.  For example, the
stellar surface density near the Sun is $\sim70\;M_\odot$ pc$^{-2}$. It
is much lower than the canonical gas threshold for star formation of
several $M_\odot$ pc$^{-2}$.  However, the outer disks of our galaxies
are gas-dominated, with gas surface densities larger than stellar by
factors of $\sim2$ to 5 or even 10 beyond several times $R_D^V$ (Hunter, private communication).
This is still below the conventional star formation threshold for gas
(see also Melena et al. 2009), but evidently not without the ability to
form stars.

A stellar surface density of $\sim0.1\;M_\odot$ pc$^{-2}$ also
corresponds to an average star formation rate of $\sim10^{-4}\;M_\odot$
yr$^{-1}$ kpc$^{-2}$ over a Gyr, which is comparable to the low end of
the star formation rates we have measured here for outer disks. It
would be 10 times lower on average if the age were a Hubble time. For a typical
galaxy rotation speed of $\sim100$ km s$^{-1}$ and a typical outer
radius of $\sim2.5$ kpc (Fig. \ref{fig-profiles}), the rotation period
is $\sim160$ Myr, so there were $\sim6$ rotations as these stars formed
in the last Gyr. Also, in a kpc wide band at 2.5 kpc radius, there is
an area of 15.7 kpc$^{2}$, so the average star formation rate in this
band over the last Gyr is $\sim10^{-3}\;M_\odot$ yr$^{-1}$. Considering
a normal IMF, where it takes $10^3\;M_\odot$ of stars overall to form a
single $\sim60\;M_\odot$ O-type star (for a Salpeter power-law slope at
intermediate to high mass and a turnover below $0.5\;M_\odot$), it
appears that the outer 1/3 of the disks of our galaxies have formed an
average of about 1 O-type star per million years.  There should,
therefore, be only a few such massive stars at any one time even if the
star formation rate were constant. 
In fact there should be far fewer in 
the cases studied here where star formation appears to have ended 
$\sim1$ Gyr ago.  A lack of O-type stars from statistically sampling a normal 
IMF at very low star formation rates could also reduce the H$\alpha$ flux below expectations.

\section{Conclusions} \label{sec-concl}

We have presented UV integrated and azimuthally-averaged surface photometric properties
of a sample of 44 dIm, BCD, and Sm galaxies. The UV measurements come from
analysis of archival NUV and FUV images obtained with {\it GALEX}, and we compare the
UV to \ha\ and $V$-band properties. We convert FUV, \ha, and $V$-band luminosities
into SFRs. We also fit model stellar populations to colors for an alternate SFR measure
for 7 different assumed star formation histories.
We compare integrated SFRs and SFR profiles with radius
in these four measures.

In most of the galaxies, the measures of SFRs track each other with 
radius. However, the UV profile often extends further in radius than 
the \ha\ profile, providing a better measure of the star formation 
activity in outer disks. Most of the dIm galaxies have constant or 
slightly increasing UV color with radius, as do the BCDs, while the Sm 
galaxies, which are also larger than the others, sometimes get bluer 
with radius. Star formation often extends so far into the outer disk that it would appear to be sub-threshold, according to conventional ideas. 
Such star formation may result from local triggering or from gravitational instabilities with angular momentum removal during cloud formation (Melena \et\ 2009).

We find that integrated SFRs determined from \ha\  are lower than SFRs
determined from FUV for all but the highest SFR systems. 
The discrepancies are unlikely
due to under-estimated extinction, plausible alternative conversion
factors, or a larger contribution from older stars where the SFR is
lower. 
On the other hand,
the SFR determined from $M_V$ is the same on average as the FUV-based
SFR for all but the lowest SFR systems, but with a large scatter about the equivalency
relationship. This is consistent with a generally constant SFR that varies by
factors of a few over long times.
In addition, the SFR
determined from modeling colors is always higher than the other 3 SFR
measures, perhaps because the star formation rate has decayed over
time, although it almost always has the same radial profile shape as $SFR_{\rm
V}$.

Galaxies with the largest discrepancies between FUV-observed star
formation rates and $V$-band observed star formation rates also have the
lowest relative HI masses. We suggest that the outer disks of these
galaxies have very low gas column densities, causing a cessation in
star formation, while in galaxies with relatively high HI masses, the
star formation in the outer disk can be active or not with fluctuations
on a Gyr timescale.

The lack of H$\alpha$ in the outer disk is most likely the result of
faint emission measures, rather than the result of radially varying
or peculiar IMFs.  This conclusion follows quantitatively from a
correlation between the ratio of star formation rates in FUV and
H$\alpha$ and the absolute H$\alpha$ star formation rate. In our
interpretation, a significant amount of H$\alpha$ flux is missing from
the outer disk star-forming regions, and so the star formation rate
determined from H$\alpha$ is too low.  We fit the observed correlation
to a model with a constant disk thickness and a density-dependent star
formation law.

The stellar surface densities in the outer parts of our galaxies reach
values as low as $\sim0.1\;M_\odot$ pc$^{-2}$, and the star formation
rates get as low as $\sim10^{-4}\;M_\odot$ yr$^{-1}$ kpc$^{-2}$ for Gyr
periods.  Most likely the star formation in these regions is
sub-threshold in the conventional sense. There is no evident break in
the FUV radial profiles from the inner disks that might correspond to a
star formation threshold, aside from the occasional presence of a kink
in the exponential disk that seems unrelated to color changes. The kink
is often far inside the furthest measured radius anyway, and thus
apparently unrelated to a physical disk edge. No such edges have been
observed yet.

\acknowledgments

BCL participated in the 2006 Research Experience for Undergraduates
(REU) program at Northern Arizona University (NAU). We appreciate Kathy
Eastwood's efforts in organizing that program and the National Science
Foundation for funding it through grant  AST-0453611 to NAU. 
Funding
for this research was provided to DAH and BGE by NASA-GALEX grant
NNX07AJ36G and by cost-sharing from Lowell Observatory. This research
has made use of the NASA/IPAC Extragalactic Database (NED) which is
operated by the Jet Propulsion Laboratory, California Institute of
Technology, under contract with the National Aeronautics and Space
Administration.
We also appreciate constructive suggestions from an anonymous referee.

Facilities: \facility{GALEX},    \facility{Lowell Observatory}

\clearpage

\clearpage

\clearpage

\clearpage

\clearpage

\plotone{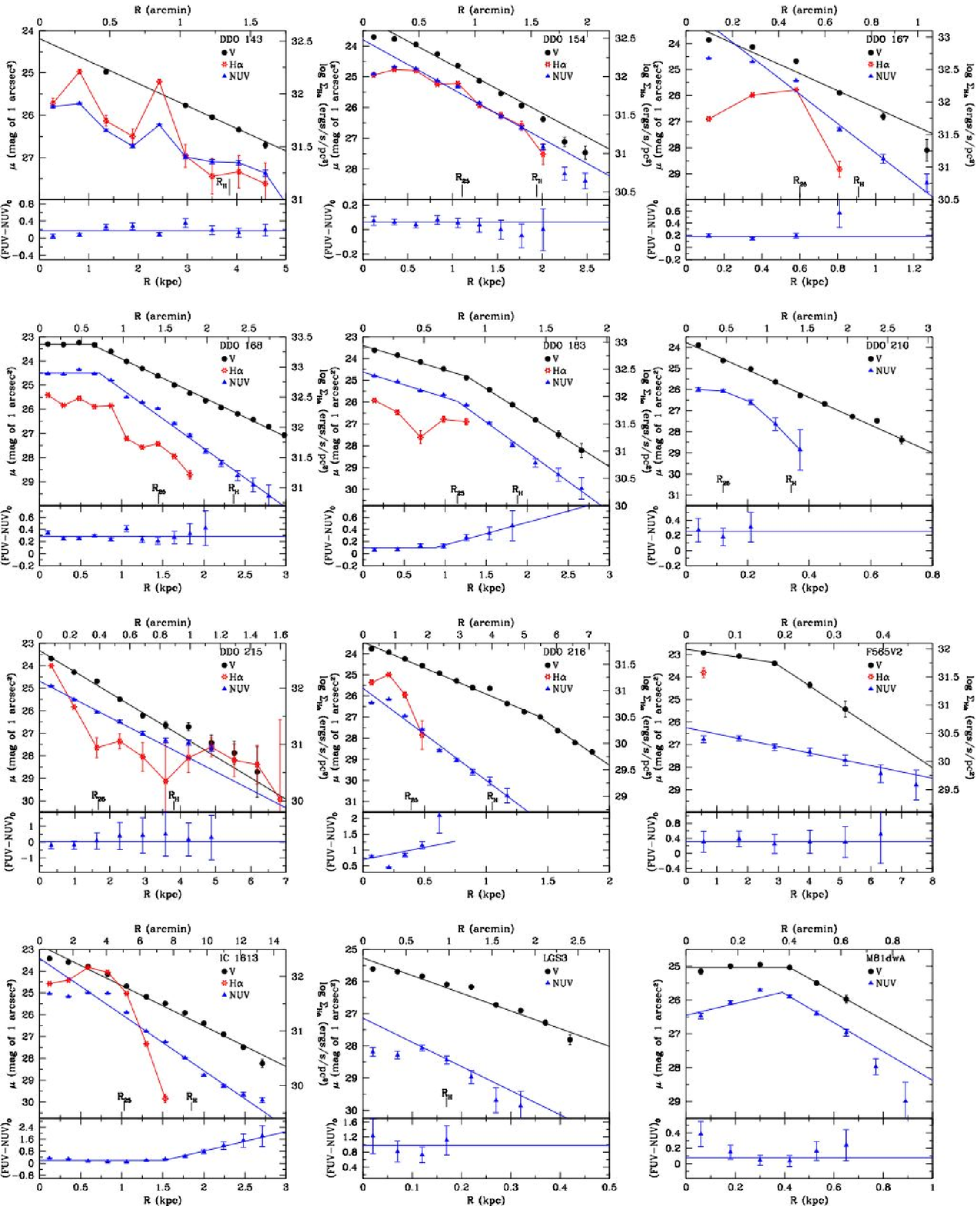}

Figure 4 (continued)

\clearpage

\plotone{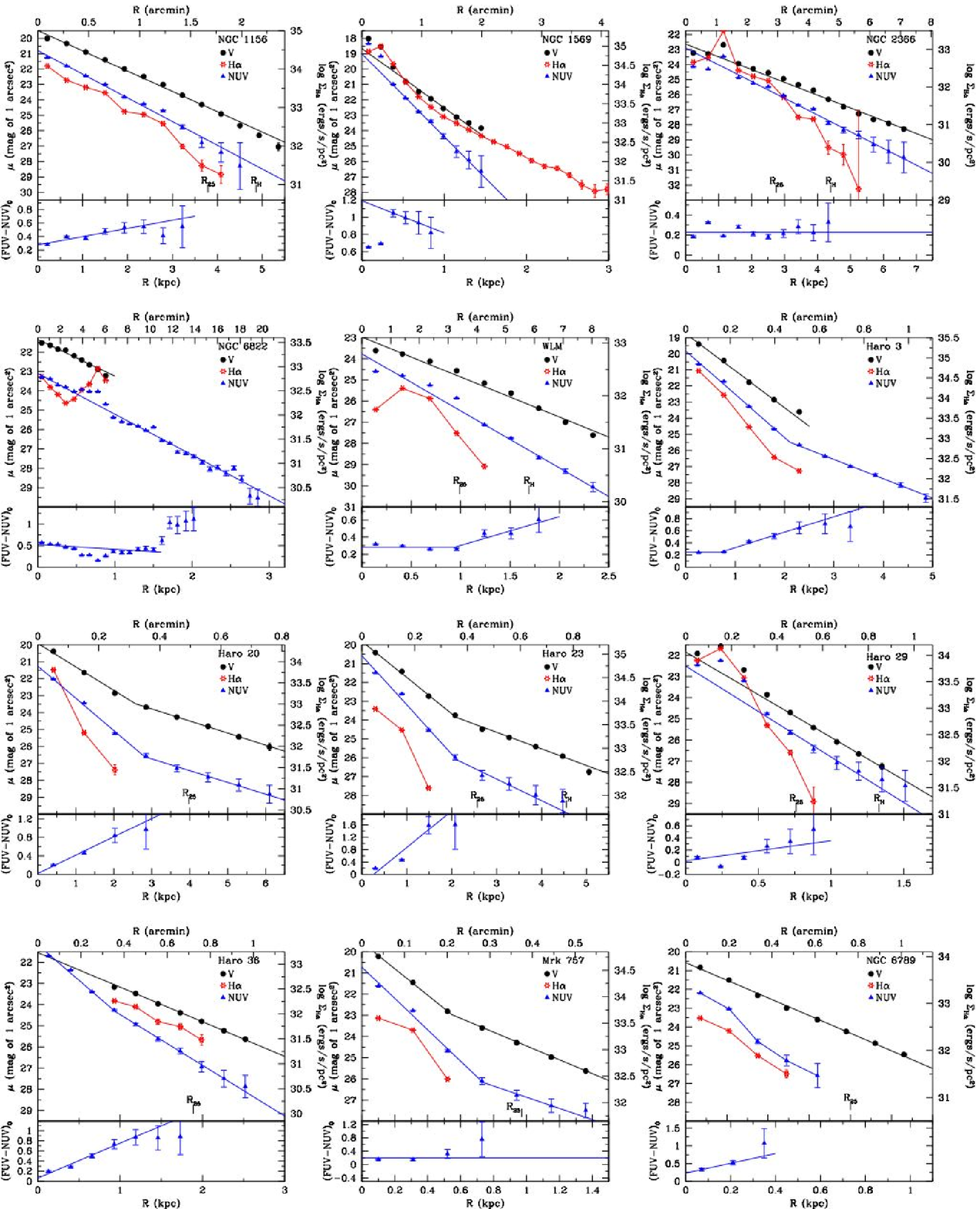}

Figure 4 (continued)

\clearpage

\plotone{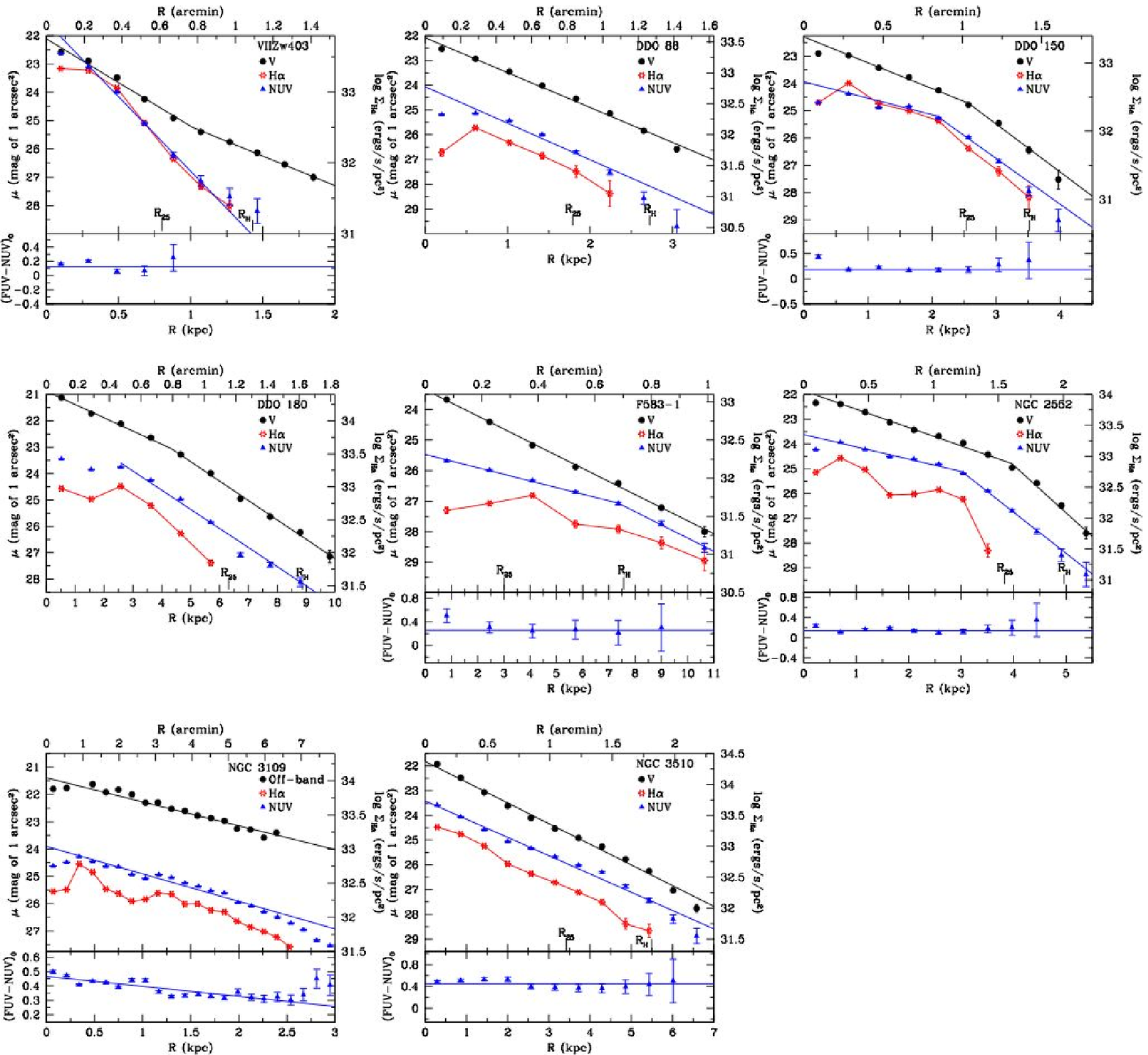}

Figure 4 (continued)

\clearpage

\clearpage

\clearpage

\clearpage

\clearpage

\clearpage

\clearpage

\clearpage

\clearpage

\clearpage

\begin{figure}
\epsscale{1.0}
\plotone{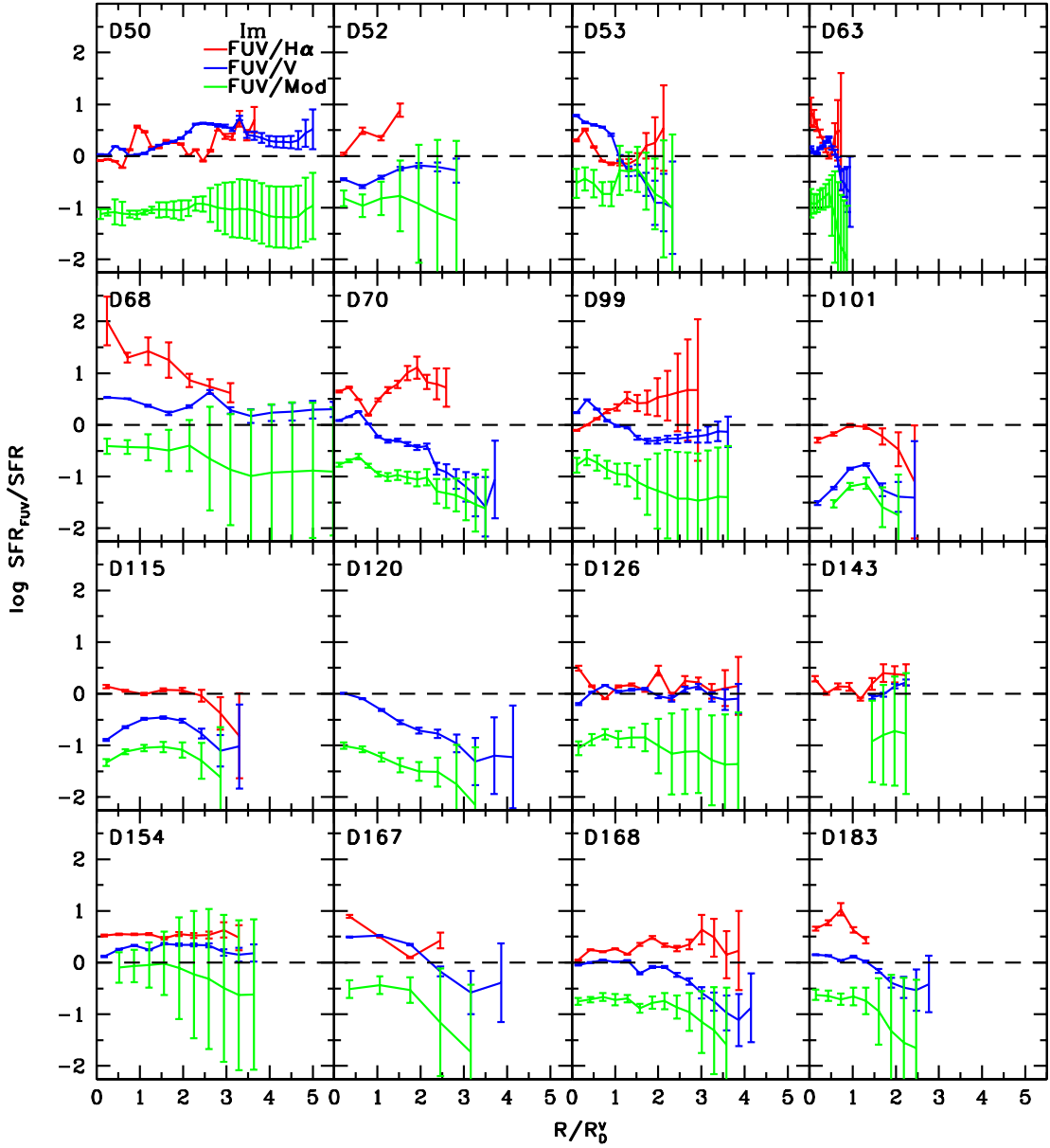}
\caption{Azimuthally-averaged star formation rates (SFRs) determined from the FUV
luminosity $SFR_{\rm FUV}$ divided by $SFR_{\rm H\alpha}$ ({\it red}),
by $SFR_{\rm V}$ ({\it blue}), and by $SFR_{\rm Col}$ ({\it green}) for the Im sample of galaxies.
$SFR_{\rm Col}$ is a SFR determined from modeling the stellar populations and star formation
histories from colors in annuli.
$SFR_{\rm V}$ is determined from the mass in stars---calculated from $M_V$ and a stellar
$M/L_V$ ratio that depends on $(B-V)_0$---and assuming a constant SFR over 12 Gyr.
The horizontal dashed lines mark equal SFRs, or ratios of 1.
The ratios are plotted as a function of the radius normalized to the disk scale length
measured from $V$ surface brightness profiles $R_D^V$.
\label{fig-sfrrat-im}}
\end{figure}

\clearpage

\plotone{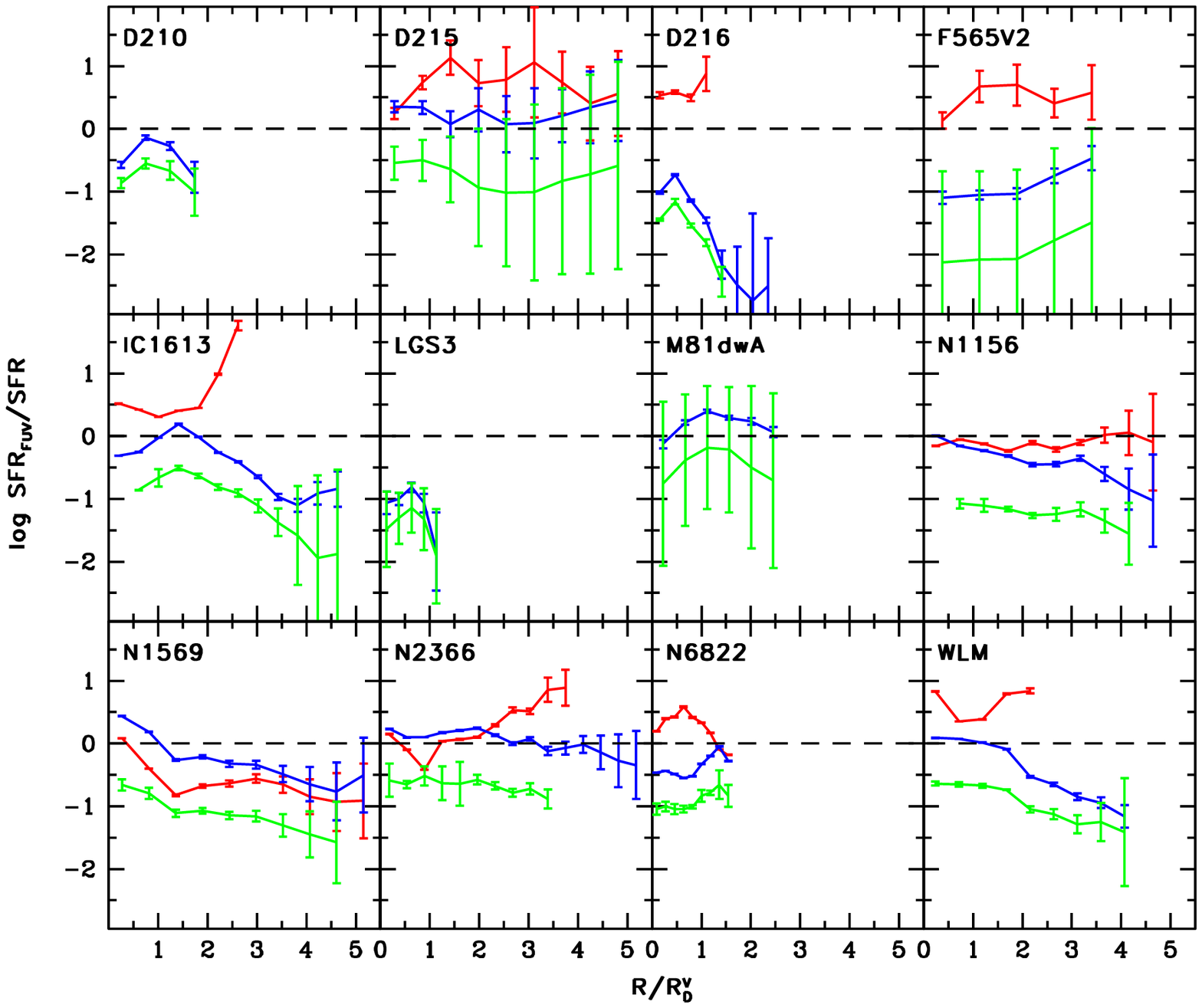}

Figure \protect\ref{fig-sfrrat-im} (continued)

\clearpage

\begin{figure}
\epsscale{1.0} \plotone{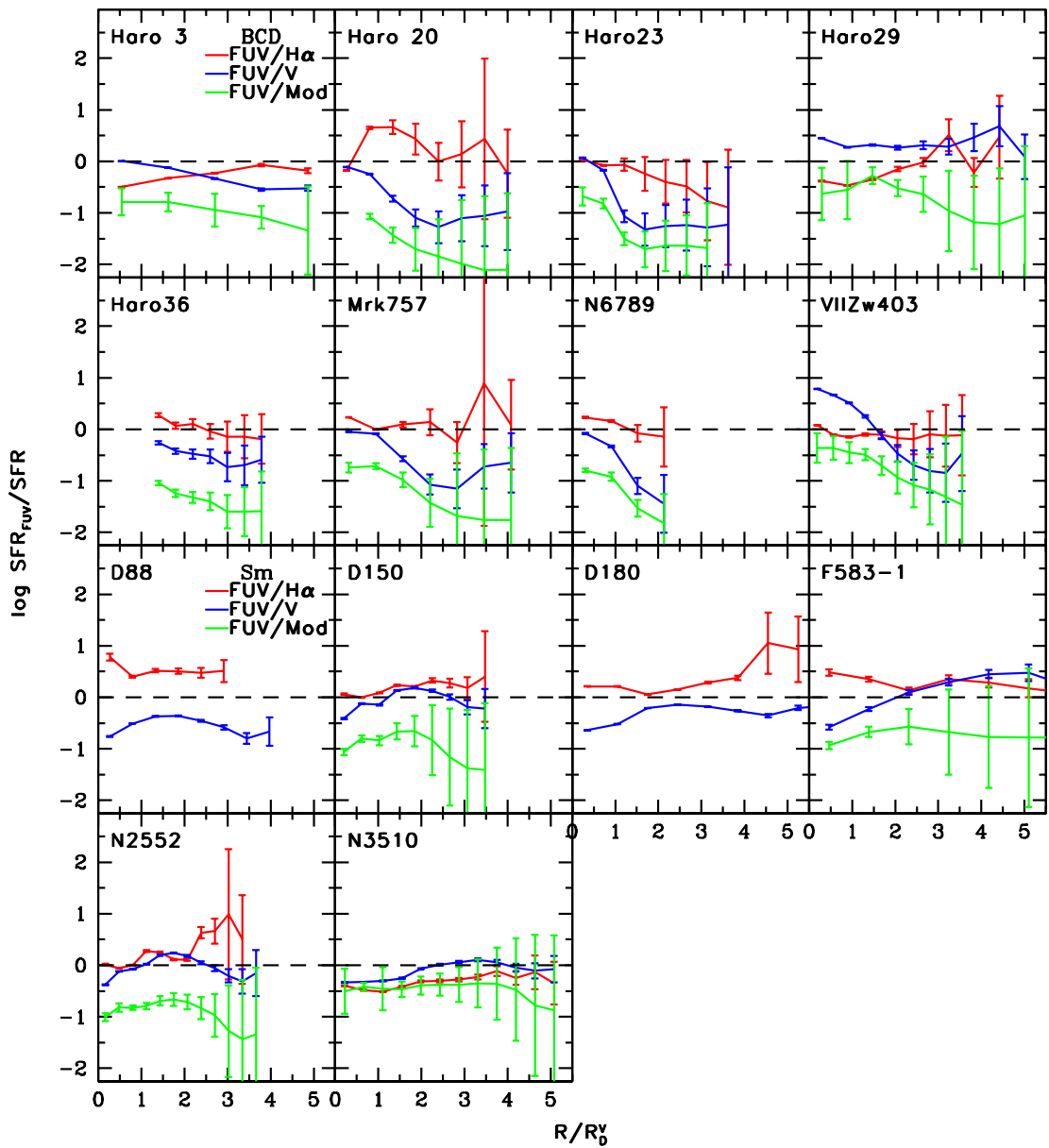} \caption{Ratios
of azimuthally-averaged of star formation rates (SFRs) versus
normalized radius, as in Fig. 14, for the BCD ({\it Top two panels})
and Sm ({\it Bottom two panels}) samples of galaxies. The BCDs tend to
have decreasing ratios with increasing radius, and the Sms tend to have
flat or increasing ratios. \label{fig-sfrrat-bcdsm}}
\end{figure}

\clearpage

\begin{figure}
\epsscale{1.0} \plotone{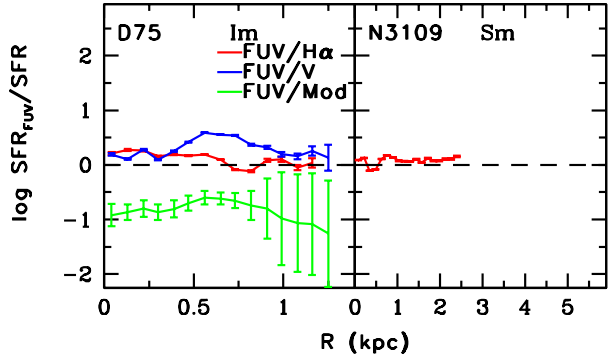}
\caption{Ratios of azimuthally-averaged star formation rates (SFRs), as
in Fig. 14, versus radius in kpc for two galaxies without $R_D^V$ measures. The
galaxy on the left is an Im system and the one on the right is an Sm.
\label{fig-sfrrat-nord}}
\end{figure}

\clearpage

\begin{figure}
\epsscale{1.0}
\plotone{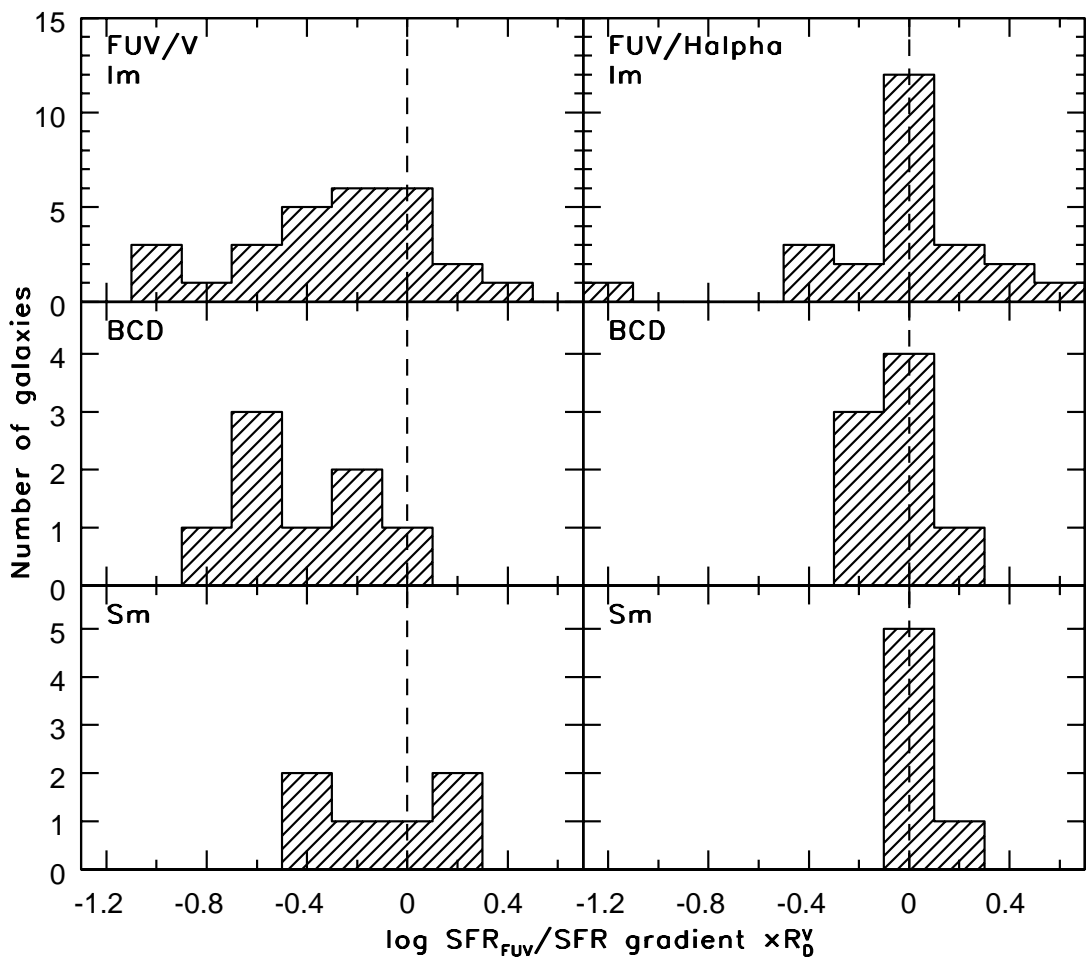}
\caption{Number of galaxies as a function of normalized SFR gradient:
$SFR_{\rm FUV}/SFR_{\rm H\alpha}$ ({\it Right}) and $SFR_{\rm FUV}/SFR_{\rm V}$ ({\it Left}).
Some of the galaxy $SFR_{\rm FUV}/SFR_{\rm V}$ profiles were fit with two parts. 
In those cases the outer gradient is counted here.
A flat radial distribution has a gradient of zero and that value is marked with a dashed
vertical line.
A positive gradient means that $SFR_{\rm FUV}$ becomes more dominant with radius;
a negative gradient means that $SFR_{\rm H\alpha}$
or $SFR_{\rm V}$ becomes more dominant with radius compared to the UV.
\label{fig-histrat}}
\end{figure}

\clearpage

\begin{figure}
\epsscale{1.0}
\plotone{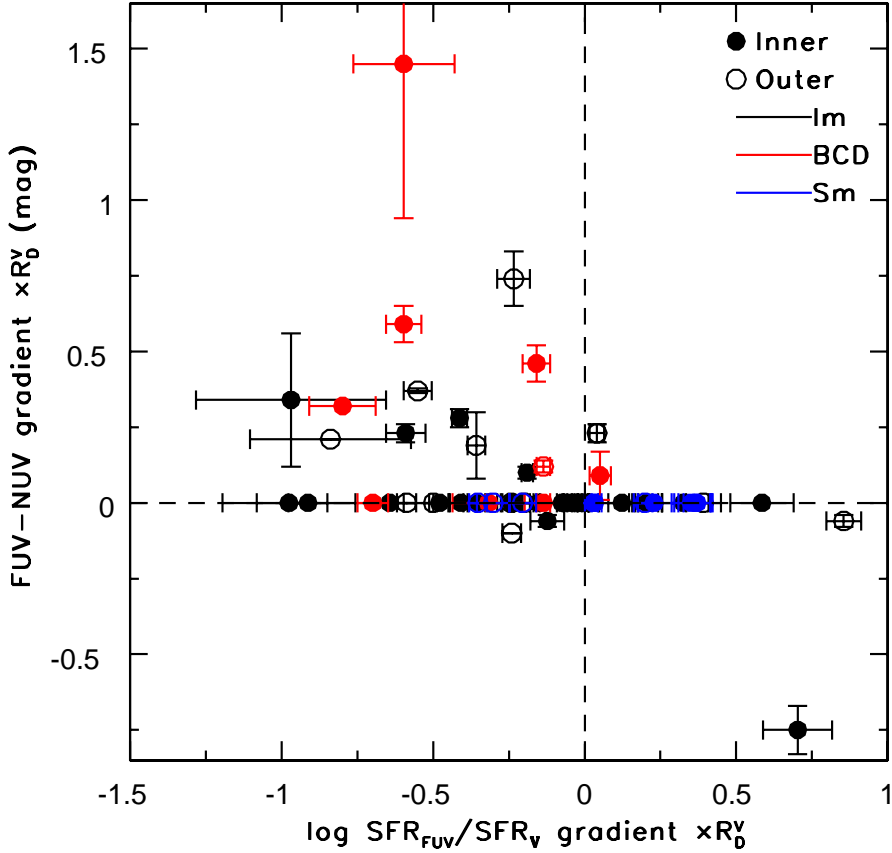}
\caption{Gradient in $(FUV-NUV)_0$ versus gradient in $\log SFR_{\rm FUV}/SFR_{\rm V}$.
Both gradients have been normalized to the $V$-band disk scale length $R_D^V$.
The UV color gradient and $SFR_{\rm FUV}/SFR_{\rm V}$ profiles, in some cases, were fit with
two components. For galaxies for which both quantities were fit with two components, the
inner pair are plotted and the outer pair of values are plotted.
For the cases where one quantity was fit with two components and the other was fit with one,
both components of the two-part fit were plotted against the
single-part fit in the other quantity.
We see that those galaxies that do show gradients in both the SFR ratio and
UV color tend to show a range in color gradients for similar SFR
ratio gradients.
Most color gradients are towards redder $(FUV-NUV)_0$ at larger radius
and towards lower $SFR_{\rm FUV}$ relative to $SFR_{\rm V}$ at larger radius.
\label{fig-gradgrad}}
\end{figure}

\clearpage

\begin{figure}
\epsscale{1.0}
\plotone{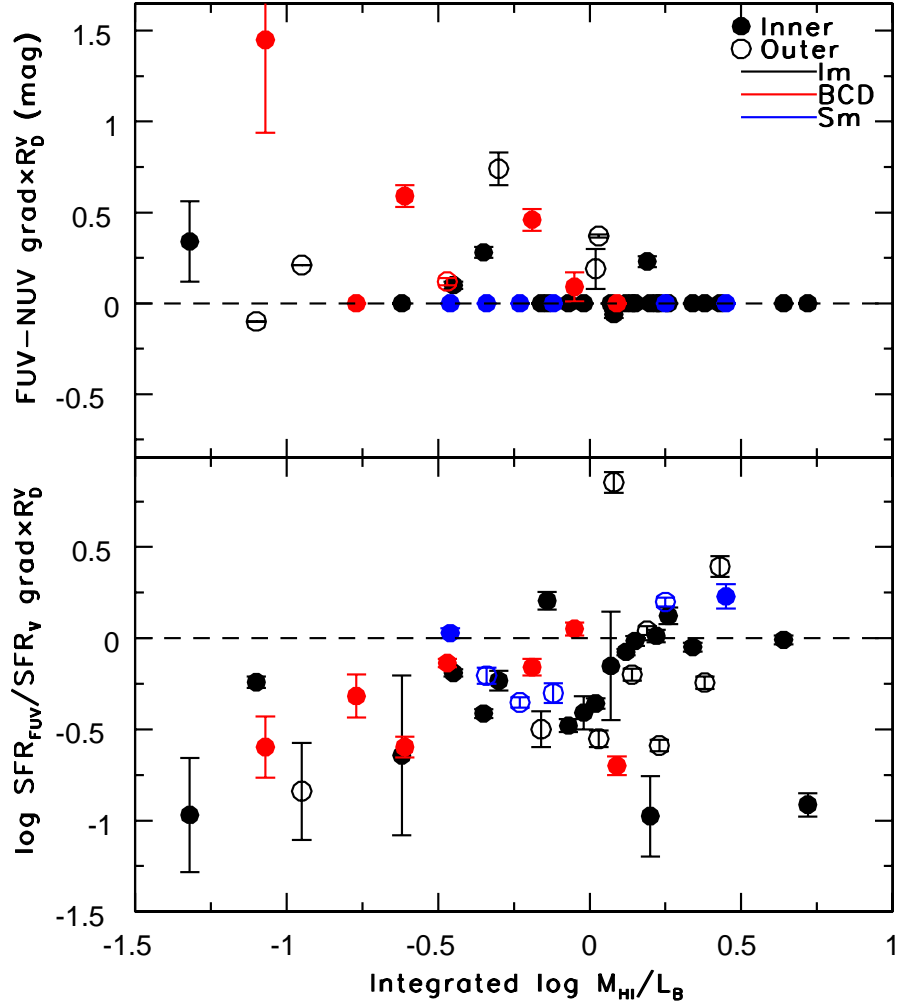}
\caption{Galactic integrated HI mass to $L_B$ ratio plotted against the gradient in
$(FUV-NUV)_0$ ({\it top)} and $\log SFR_{\rm FUV}/SFR_{\rm V}$ ({\it bottom}).
Both gradients have been normalized to the $V$-band disk scale length $R_D^V$.
The horizontal dashed lines mark gradients of zero.
For profiles fit with two lines, both the inner and outer gradients are plotted.
\label{fig-normalizemhilb2}}
\end{figure}

\clearpage

\begin{figure}
\epsscale{1.0}
\plotone{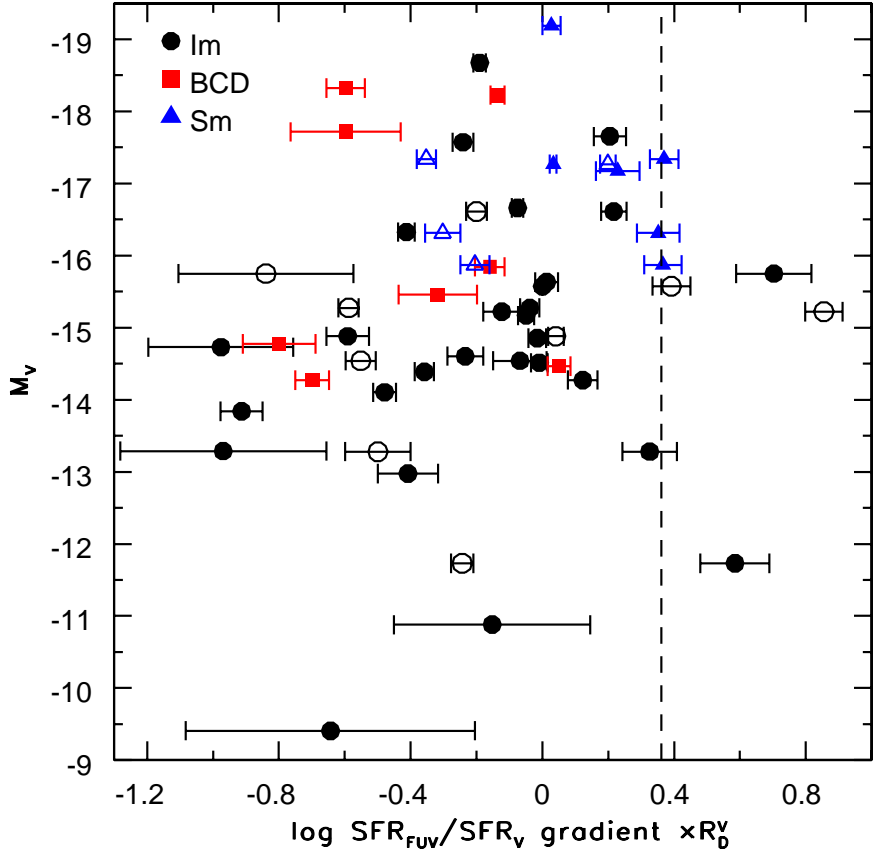}
\caption{Gradient in $\log SFR_{\rm FUV}/SFR_{\rm V}$
normalized to the $V$-band disk scale length $R_D^V$ vs. galactic $M_V$.
Filled symbols are for inner $SFR_{\rm FUV}/SFR_{\rm V}$ gradients
and open symbols are for outer $SFR_{\rm FUV}/SFR_{\rm V}$ gradients 
for those profiles that were fit with two lines.
\label{fig-gradmv}}
\end{figure}

\clearpage

\begin{figure}
\epsscale{1.0} 
\plotone{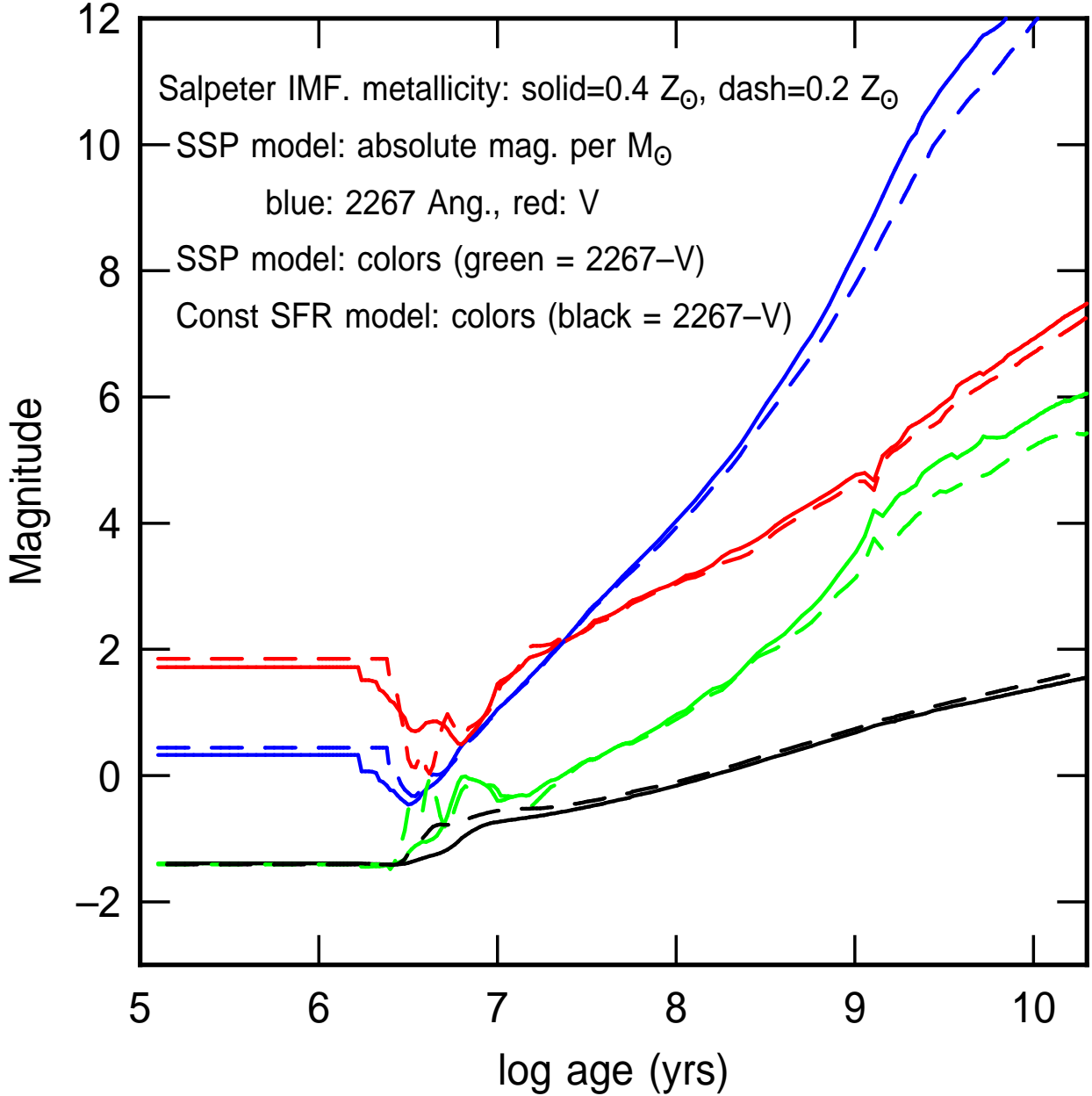} 
\caption{Models from Bruzual \&
Charlot (2003) of the evolution of absolute magnitude for a single
population of $1\;M_\odot$ of stars at 2267 \AA\ and $V$ band ({\it blue} and
{\it red} lines, respectively). {\it Green} curves show the $2267-V$ colors in that
model. The black curves are for a constant star formation rate. Because
the observations suggest red $2267-V$ colors of 1 to 3 magnitudes in the
outer parts of our sample disks, only the SSP models with star
formation ending $\sim100$ Myr ago or longer are acceptable. Two
metallicities are represented by different line types.
{\it GALEX} magnitudes in the rest of this paper are on the AB scale and $V$ 
magnitudes are on the Vega scale. Bruzual \& Charlot (2003) magnitudes 
are also on the Vega scale. To make the color $NUV-V$ in this figure 
have the same meaning as in the rest of this paper, we added 1.8 to the 
Bruzual \& Charlot NUV magnitudes to put them on the AB scale. This 
correction comes from the Vega flux (Castelli \& Kurucz 1994) for the 
effective wavelength of the NUV filter, 2267\AA.
\label{fig-mag22}}
\end{figure}

\clearpage

\begin{figure}
\epsscale{1.0} 
\plotone{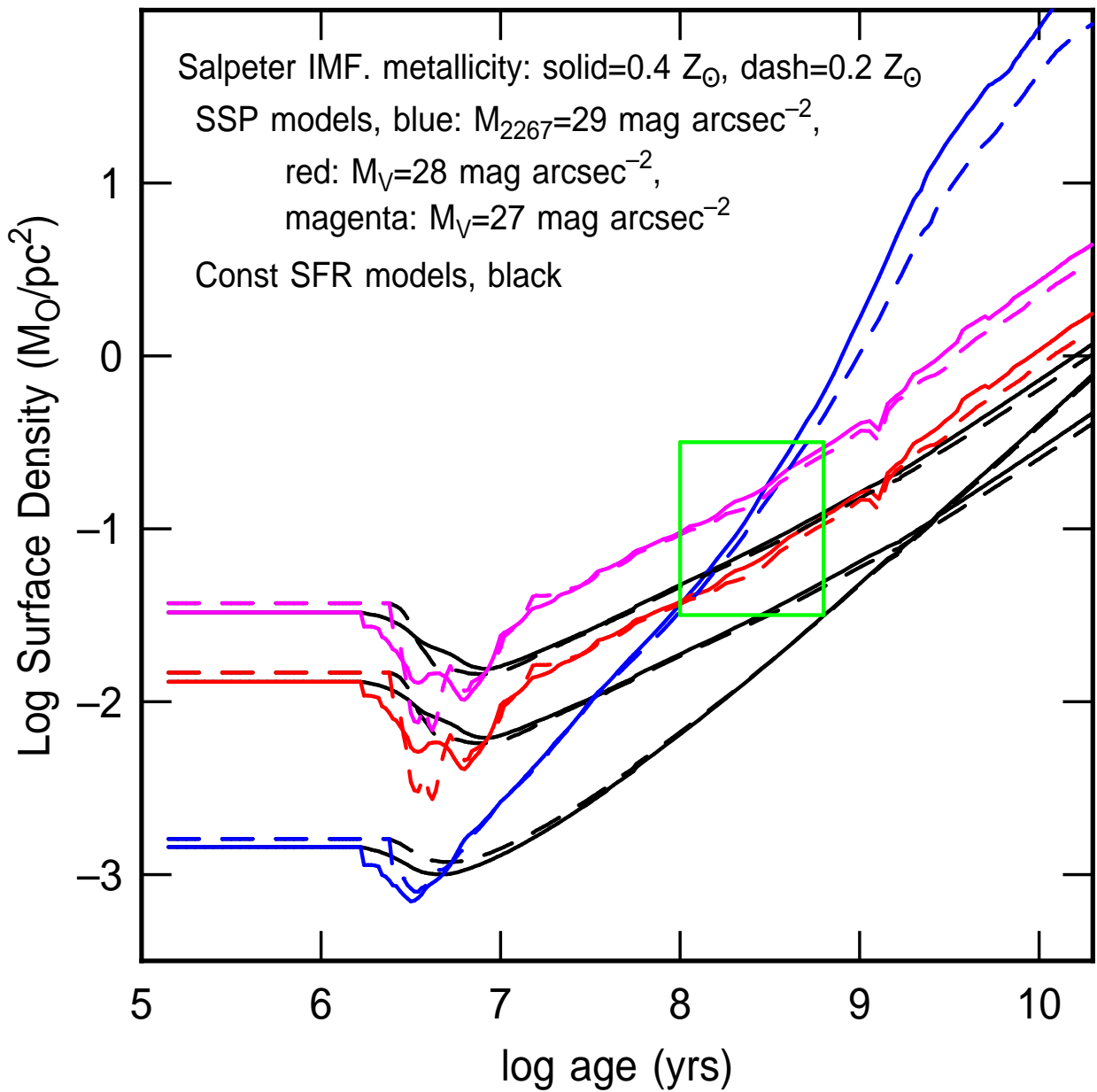} 
\caption{Mass surface density
is shown versus population age for a surface brightness at 2267 \AA\ of
29 mag arcsec$^{-2}$ (blue curves) and surface brightnesses at $V$-band
of 27 and 28 mag arcsec$^{-2}$ (magenta and red, respectively).  The 2267 \AA\ and
$V$-band curves intersect at the most likely value for the observations,
which is indicated by the {\it green} box. 
This result suggests that the outer disk has an age of several hundred 
Myr and a surface density of
$\sim0.1\;M_\odot$ pc$^{-2}$.  The {\it black} lines are for a constant star
formation rate, which does not give a solution 
(i.e., the same surface density fit for both passbands) 
in a reasonable age.
Two metallicities are represented by different line types.
\label{fig-mag22-sigma}}
\end{figure}

\clearpage

                                                                           
%
                                                                           
\begin{deluxetable}{llrcrrl}                                               
\tabletypesize{\scriptsize}                                                
\rotate                                                                    
\tablenum{1}                                                               
\tablecolumns{7}                                                           
\tablewidth{0pt}                                                           
\tablecaption{Galaxy Sample and {\it GALEX} Observations\label{tab-genobs}}
\tablehead{                                                                
\colhead{}                                                                 
& \colhead{}                                                               
& \colhead{D\tablenotemark{a}}                                             
& \colhead{}                                                               
& \colhead{}                                                               
& \colhead{}                                                               
& \colhead{} \\                                                            
\colhead{Galaxy}                                                           
& \colhead{Other Names}                                                    
& \colhead{(Mpc)}                                                          
& \colhead{E($B-V$)$_f$\tablenotemark{b}}                                  
& \colhead{FUV Exp (s)}                                                    
& \colhead{NUV Exp (s)}                                                    
& \colhead{Tile Name}                                                      
}                                                                          
\startdata                                                                 
\cutinhead{Im Galaxies}                                                    
DDO 50  \dotfill & PGC 23324,UGC 4305,Holmberg II,VIIZw223    &  3.4 & 0.02 &  1674\xxx &  1521\xxx & NGA\_HolmbergII             \\
DDO 52  \dotfill & PGC 23769,UGC 4426                         &  6.0 & 0.03 &  1678\xxx &  1522\xxx & GI1\_047032\_UGC04426\_0001 \\
DDO 53  \dotfill & PGC 24050,UGC 4459,VIIZw238                &  3.6 & 0.03 &  1672\xxx &  1502\xxx & NGA\_DDO053                 \\
DDO 63  \dotfill & PGC 27605,Holmberg I,UGC 5139              &  3.8 & 0.01 &  1676\xxx &  1544\xxx & NGA\_HolmbergI              \\
DDO 68  \dotfill & PGC 28714,UGC 5340                         &  7.0 & 0.01 &  1435\xxx &  1305\xxx & GI1\_047036\_UGC05340       \\
DDO 70  \dotfill & PGC 28913,UGC 5373,Sextans B               &  1.3 & 0.01 &  1076\xxx &   966\xxx & NGA\_SextansB               \\
DDO 75  \dotfill & PGC 29653,UGCA 205,Sextans A               &  1.3 & 0.02 &  1663\xxx &  1512\xxx & NGA\_SextansA               \\
DDO 99  \dotfill & PGC 37050,UGC 6817                         &  4.1 & 0.00 &  2348\xxx &  2196\xxx & GI1\_047053\_UGC06817       \\
\enddata                                                                   
\tablenotetext{a}{See references in Hunter \& Elmegreen (2006).}           
\tablenotetext{b}{Foreground reddening from Burstein \& Heiles (1984).}    
\end{deluxetable}                                                          

\clearpage

                                                                                         
%
                                                                                         
\begin{deluxetable}{lcccccccccccc}                                                       
\tabletypesize{\scriptsize}                                                              
\rotate                                                                                  
\tablenum{2}                                                                             
\tablecolumns{13}                                                                        
\tablewidth{0pt}                                                                         
\tablecaption{Integrated Optical and Infrared Photometry\tablenotemark{a}\label{tab-opt}}
\tablehead{                                                                              
\colhead{}                                                                               
& \colhead{}                                                                             
& \colhead{}                                                                             
& \colhead{}                                                                             
& \colhead{}                                                                             
& \colhead{}                                                                             
& \colhead{}                                                                             
& \colhead{}                                                                             
& \colhead{}                                                                             
& \colhead{}                                                                             
& \colhead{}                                                                             
& \colhead{log L$_{H\alpha}$}                                                            
& \colhead{$\sigma_{\log L_{H\alpha}}$} \\                                               
\colhead{Galaxy}                                                                         
& \colhead{M$_V$}                                                                        
& \colhead{$\sigma_{M_V}$}                                                               
& \colhead{($U-B$)$_0$}                                                                  
& \colhead{$\sigma_{U-B}$}                                                               
& \colhead{($B-V$)$_0$}                                                                  
& \colhead{$\sigma_{B-V}$}                                                               
& \colhead{($J-H$)$_0$}                                                                  
& \colhead{$\sigma_{J-H}$}                                                               
& \colhead{($H-K$)$_0$}                                                                  
& \colhead{$\sigma_{H-K}$}                                                               
& \colhead{(ergs s$^{-1}$)}                                                              
& \colhead{(ergs s$^{-1}$)}                                                              
}                                                                                        
\startdata                                                                               
\cutinhead{Im Galaxies}                                                                  
DDO 50  \dotfill & -16.61 & 0.00 & -0.41 & 0.00 &  0.22 & 0.00 & \nodata & \nodata & \nodata & \nodata & 40.025 & 0.001 \\
DDO 52  \dotfill & -14.27 & 0.03 & -0.02 & 0.10 &  0.39 & 0.04 & \nodata & \nodata & \nodata & \nodata & 38.213 & 0.013 \\
DDO 53  \dotfill & -13.84 & 0.01 & -0.53 & 0.01 &  0.41 & 0.01 &  0.45 & 2.17 & \nodata & \nodata & 38.947 & 0.001 \\
DDO 63  \dotfill & -14.73 & 0.02 & -0.18 & 0.05 &  0.20 & 0.02 & \nodata & \nodata & \nodata & \nodata & 38.977 & 0.010 \\
DDO 68  \dotfill & -15.17 & 0.02 & -0.24 & 0.05 &  0.12 & 0.03 & \nodata & \nodata & \nodata & \nodata & 38.457 & 0.040 \\
DDO 70  \dotfill & -14.10 & 0.00 & -0.11 & 0.00 &  0.36 & 0.00 & \nodata & \nodata & \nodata & \nodata & 38.171 & 0.003 \\
DDO 75  \dotfill & -13.91 & 0.01 & -0.54 & 0.02 &  0.19 & 0.01 & \nodata & \nodata & \nodata & \nodata & 39.060 & 0.001 \\
DDO 99  \dotfill & -14.88 & 0.01 & -0.20 & 0.03 &  0.28 & 0.02 & \nodata & \nodata & \nodata & \nodata & 38.931 & 0.006 \\
\enddata                                                                                 
\tablenotetext{a}{From Hunter \& Elmegreen (2004, 2006).}                                
\end{deluxetable}                                                                        

\clearpage

                                                                         
%
                                                                         
\begin{deluxetable}{lrcccccccc}                                          
\tabletypesize{\scriptsize}                                              
\rotate                                                                  
\tablenum{3}                                                             
\tablecolumns{10}                                                        
\tablewidth{0pt}                                                         
\tablecaption{Integrated {\it GALEX} Photometry\label{tab-phot}}         
\tablehead{                                                              
\colhead{Galaxy}                                                         
& \colhead{M$_{NUV}$\tablenotemark{a}}                                   
& \colhead{$\sigma_{M_{NUV}}$}                                           
& \colhead{($FUV-NUV$)$_0$}                                              
& \colhead{$\sigma_{FUV-NUV}$}                                           
& \colhead{($NUV-V$)$_0$\tablenotemark{a}}                               
& \colhead{$\sigma_{NUV-V}$}                                             
& \colhead{$\log L_{H\alpha}/L_{NUV}$\tablenotemark{b}}                  
& \colhead{$\log L_{NUV}$\tablenotemark{c}}                              
& \colhead{$\sigma_{log L_{NUV}}$}                                       
}                                                                        
\startdata                                                               
\cutinhead{Im Galaxies}                                                  
DDO 50  \dotfill & -15.77 & 0.00 &  0.08 & 0.00 & 0.88 & 0.00 & -1.59 & 26.98 & 0.00 \\
DDO 52  \dotfill & -12.65 & 0.01 &  0.47 & 0.02 & 1.62 & 0.04 & -2.17 & 25.73 & 0.00 \\
DDO 53  \dotfill & -12.76 & 0.01 &  0.15 & 0.02 & 1.08 & 0.01 & -1.48 & 25.80 & 0.00 \\
DDO 63  \dotfill & -13.44 & 0.00 &  0.23 & 0.01 & 1.29 & 0.02 & -1.72 & 26.04 & 0.00 \\
DDO 68  \dotfill & -14.43 & 0.00 &  0.14 & 0.01 & 0.74 & 0.02 & -2.64 & 26.44 & 0.00 \\
DDO 70  \dotfill & -12.41 & 0.00 &  0.38 & 0.01 & 1.69 & 0.00 & -2.12 & 25.63 & 0.00 \\
DDO 75  \dotfill & -13.26 & 0.00 &  0.21 & 0.00 & 0.66 & 0.01 & -1.57 & 25.97 & 0.00 \\
DDO 99  \dotfill & -13.62 & 0.00 &  0.34 & 0.01 & 1.26 & 0.01 & -1.84 & 26.12 & 0.00 \\
\enddata                                                                 
\tablenotetext{a}{$M_{NUV}$ is an AB magnitude.}                         
\tablenotetext{b}{$L_{NUV}$ here is in ergs s$^{-1}$,                    
as is $L_{H\alpha}$. $L_{NUV}$ has been converted to ergs s$^{-1}$       
by multiplying the flux in ergs s$^{-1}$ cm$^{-2}$ by the FWHM of        
the bandpass of the $NUV$ filter in Angstroms.}                          
\tablenotetext{c}{From $f_{\nu}$ in ergs s$^{-1}$ Hz$^{-1}$.}            
\end{deluxetable}                                                        

\clearpage

                                                                         
%
                                                                         
\begin{deluxetable}{lccrrrrrccrrrr}                                      
\tabletypesize{\scriptsize}                                              
\rotate                                                                  
\tablenum{4}                                                             
\tablecolumns{14}                                                        
\tablewidth{0pt}                                                         
\tablecaption{Disk Structure\label{tab-struc}}                           
\tablehead{                                                              
\colhead{}                                                               
& \multicolumn{6}{c}{Inner}                                              
& \colhead{}                                                             
& \multicolumn{6}{c}{Outer} \\                                           
\cline{2-7}                                                              
\cline{9-14}                                                             
\colhead{}                                                               
& \colhead{$\mu_0^{NUV}$}                                                
& \colhead{$\sigma_{\mu_0^{NUV}}$}                                       
& \colhead{$R_D^{NUV}$\tablenotemark{a}}                                 
& \colhead{$\sigma_{R_D^{NUV}}$}                                         
& \colhead{$R_D^V$\tablenotemark{a}}                                     
& \colhead{$\sigma_{R_D^V}$}                                             
& \colhead{}                                                             
& \colhead{$\mu_0^{NUV}$}                                                
& \colhead{$\sigma_{\mu_0^{NUV}}$}                                       
& \colhead{$R_D^{NUV}$\tablenotemark{a}}                                 
& \colhead{$\sigma_{R_D^{NUV}}$}                                         
& \colhead{$R_D^V$\tablenotemark{a}}                                     
& \colhead{$\sigma_{R_D^V}$} \\                                          
\colhead{Galaxy}                                                         
& \colhead{(mag arcsec$^{-2}$)}                                          
& \colhead{(mag arcsec$^{-2}$)}                                          
& \colhead{(kpc)}                                                        
& \colhead{(kpc)}                                                        
& \colhead{(kpc)}                                                        
& \colhead{(kpc)}                                                        
& \colhead{}                                                             
& \colhead{(mag arcsec$^{-2}$)}                                          
& \colhead{(mag arcsec$^{-2}$)}                                          
& \colhead{(kpc)}                                                        
& \colhead{(kpc)}                                                        
& \colhead{(kpc)}                                                        
& \colhead{(kpc)}                                                        
}                                                                        
\startdata                                                               
\cutinhead{Im Galaxies}                                                  
DDO 50  \dotfill & \nodata & \nodata & \nodata & \nodata &  1.10 & 0.05 & & \nodata & \nodata & \nodata & \nodata & \nodata & \nodata \\
DDO 52  \dotfill & 25.19 & 0.09 &  0.86 & 0.05 &  0.76 & 0.02 & & \nodata & \nodata & \nodata & \nodata & \nodata & \nodata \\
DDO 53  \dotfill & 23.05 & 0.24 &  0.30 & 0.01 &  0.72 & 0.06 & & \nodata & \nodata & \nodata & \nodata & \nodata & \nodata \\
DDO 63  \dotfill & 25.62 & 0.03 & \nodata & \nodata &  3.01 & 0.49 & & 21.63 & 0.12 &  0.37 & 0.01 &  0.66 & 0.01 \\
DDO 68  \dotfill & 23.21 & 0.17 &  0.62 & 0.05 &  0.60 & 0.02 & & 26.67 & 0.24 &  3.75 & 1.12 &  2.97 & 3.28 \\
DDO 70  \dotfill & 24.04 & 0.10 &  0.35 & 0.01 &  0.47 & 0.00 & & \nodata & \nodata & \nodata & \nodata & \nodata & \nodata \\
DDO 75  \dotfill & 24.93 & 0.07 & \nodata & \nodata & \nodata & \nodata & & 20.34 & 0.20 &  0.17 & 0.01 &  0.22 & 0.01 \\
DDO 99  \dotfill & 24.35 & 0.06 &  0.72 & 0.02 &  0.97 & 0.02 & & \nodata & \nodata & \nodata & \nodata &  0.64 & 0.02 \\
\enddata                                                                 
\tablenotetext{a}{$R_D$ are disk scale lengths. The                      
disk scale lengths determined from the $V$-band $R_D^V$                  
are from Hunter \& Elmegreen (2006).}                                    
\end{deluxetable}                                                        

\clearpage

                                                                         
%
                                                                         
\begin{deluxetable}{lccrrrrrcc}                                      
\tabletypesize{\scriptsize}                                              
\rotate                                                                  
\tablenum{5}                                                             
\tablecolumns{10}                                                        
\tablewidth{0pt}                                                         
\tablecaption{$FUV-NUV$ Color Gradients\label{tab-colorgrad}}                           
\tablehead{                                                              
\colhead{}                                                               
& \multicolumn{4}{c}{Inner}                                              
& \colhead{}                                                             
& \multicolumn{4}{c}{Outer} \\                                           
\cline{2-5}                                                              
\cline{7-10}                                                             
\colhead{}                                                               
& \colhead{}                                          
& \colhead{}                                   
& \colhead{gradient\tablenotemark{b}}                                 
& \colhead{$\sigma_{\rm gradient}$}                                         
& \colhead{$R_{\rm transition}$\tablenotemark{c}}                                                           
& \colhead{}                                                             
& \colhead{}                                             
& \colhead{gradient\tablenotemark{b}}                                 
& \colhead{$\sigma_{\rm gradient}$}      \\             
\colhead{Galaxy}   
& \colhead{$(FUV-NUV)^0$\tablenotemark{a}}                                                
& \colhead{$\sigma_{(FUV-NUV)^0}$}                                                                                  
& \colhead{(mag kpc$^{-1}$)}                                                        
& \colhead{(mag kpc$^{-1}$)}                                                        
& \colhead{(kpc)}  
& \colhead{$(FUV-NUV)^0$\tablenotemark{a}}                                                
& \colhead{$\sigma_{(FUV-NUV)^0}$}                                                                                         
& \colhead{(mag kpc$^{-1}$)}                                                        
& \colhead{(mag kpc$^{-1}$)}                                       
}                                                                        
\startdata                                                               
\cutinhead{Im Galaxies}                                                  
DDO 50  \dotfill & 0.10 & 0.08 & \nodata & \nodata & \nodata & \nodata  & \nodata  & \nodata & \nodata \\
DDO 52  \dotfill & 0.46 & 0.07 & \nodata & \nodata & \nodata & \nodata  & \nodata & \nodata & \nodata  \\
DDO 53  \dotfill & 0.03 & 0.07 & \nodata & \nodata & \nodata & \nodata  & \nodata & \nodata & \nodata  \\
DDO 63  \dotfill & 0.20 & 0.07 & \nodata & \nodata & \nodata & \nodata  & \nodata & \nodata & \nodata \\
DDO 68  \dotfill & 0.15 & 0.06 & \nodata & \nodata & \nodata & \nodata  & \nodata & \nodata & \nodata  \\
DDO 70  \dotfill & 0.42 & 0.05 & \nodata & \nodata & \nodata & \nodata  & \nodata & \nodata & \nodata  \\
DDO 75  \dotfill & 0.43 & 0.02 & $-$0.38 & 0.04 & \nodata & \nodata  & \nodata & \nodata & \nodata  \\
DDO 99  \dotfill & 0.09 & 0.04 &  0.24 &  0.03 & \nodata & \nodata & \nodata & \nodata & \nodata  \\
\enddata                                                                 
\tablenotetext{a}{Central color. An entry here
and no entry under the gradient means that the color is constant.}
\tablenotetext{b}{$(FUV-NUV)_0$ color gradient. 
A color gradient without an uncertainty means that the central color and gradient
were determined from only two points.} 
\tablenotetext{b}{For profiles that are fit in two parts, this is the radius where the two
fits cross.}
\end{deluxetable}                                                        

\clearpage

                                                                         
%
                                                                         
\begin{deluxetable}{lccccccccccc}                                        
\tabletypesize{\scriptsize}                                              
\rotate                                                                  
\tablenum{6}                                                             
\tablecolumns{12}                                                        
\tablewidth{0pt}                                                         
\tablecaption{Integrated Star Formation Rates\label{tab-sfr}}            
\tablehead{                                                              
\colhead{}                                                               
& \colhead{log SFR$_{FUV}$\tablenotemark{a}}                             
& \colhead{$\sigma$\tablenotemark{b}}                                    
& \colhead{log SFR$_{H\alpha}$}                                          
& \colhead{$\sigma$\tablenotemark{b}}                                    
& \colhead{log SFR$_{V}$}                                                
& \colhead{$\sigma$\tablenotemark{b}}                                    
& \colhead{$M/L_V$}                                                      
& \colhead{}                                                             
& \colhead{}                                                             
& \colhead{}                                                             
& \colhead{} \\                                                          
\colhead{Galaxy}                                                         
& \colhead{($M_\sun$ yr$^{-1}$)}                                         
& \colhead{($M_\sun$ yr$^{-1}$)}                                         
& \colhead{($M_\sun$ yr$^{-1}$)}                                         
& \colhead{($M_\sun$ yr$^{-1}$)}                                         
& \colhead{($M_\sun$ yr$^{-1}$)}                                         
& \colhead{($M_\sun$ yr$^{-1}$)}                                         
& \colhead{($M_\sun/L_{V,\sun}$)}                                        
& \colhead{log SFR$_{FUV}$/SFR$_{H\alpha}$}                              
& \colhead{$\sigma$\tablenotemark{b}}                                    
& \colhead{log SFR$_{FUV}$/SFR$_{V}$}                                    
& \colhead{$\sigma$\tablenotemark{b}}                                    
}                                                                        
\startdata                                                               
\cutinhead{Im Galaxies}                                                  
DDO 50  \dotfill &  -0.98 &  0.00 & -1.14 & 0.00 & -1.32 & 0.00 & 0.75 &  0.16 & 0.00 &  0.34 & 0.00 \\
DDO 52  \dotfill &  -2.38 &  0.01 & -2.95 & 0.01 & -2.06 & 0.01 & 1.20 &  0.56 & 0.01 &  0.33 & 0.01 \\
DDO 53  \dotfill &  -2.20 &  0.01 & -2.21 & 0.00 & -2.21 & 0.00 & 1.25 &  0.01 & 0.01 &  0.01 & 0.01 \\
DDO 63  \dotfill &  -1.97 &  0.00 & -2.18 & 0.01 & -2.10 & 0.01 & 0.71 &  0.21 & 0.01 &  0.12 & 0.01 \\
DDO 68  \dotfill &  -1.54 &  0.00 & -2.74 & 0.08 & -2.01 & 0.01 & 0.58 &  1.21 & 0.08 &  0.47 & 0.01 \\
DDO 70  \dotfill &  -2.44 &  0.00 & -2.99 & 0.00 & -2.16 & 0.00 & 1.10 &  0.55 & 0.00 &  0.28 & 0.00 \\
DDO 75  \dotfill &  -2.04 &  0.00 & -2.10 & 0.00 & -2.43 & 0.00 & 0.70 &  0.06 & 0.00 &  0.39 & 0.00 \\
DDO 99  \dotfill &  -1.93 &  0.00 & -2.20 & 0.01 & -1.94 & 0.01 & 0.90 &  0.35 & 0.01 &  0.08 & 0.01 \\
\enddata                                                                 
\tablenotetext{a}{Determined from FUV.}                                  
\tablenotetext{b}{The $\sigma$ refers to the uncertainty in the column   
to the left.}                                                            
\end{deluxetable}                                                        

\clearpage

                                                                                 
%
                                                                                 
\begin{deluxetable}{lccccccccccccccc}                                            
\tabletypesize{\scriptsize}                                                      
\rotate                                                                          
\tablenum{7}                                                                     
\tablecolumns{16}                                                                
\tablewidth{0pt}                                                                 
\tablecaption{Model Fits to Integrated Colors\label{tab-intmod}}                 
\tablehead{                                                                      
\colhead{}                                                                       
& \colhead{}                                                                     
& \colhead{$\log$ Decay t\tablenotemark{a}}                                      
& \colhead{$\log$ Age}                                                           
& \colhead{}                                                                     
& \colhead{$\log$ Mass}                                                          
& \colhead{}                                                                     
& \colhead{$\log SFR_{\rm Col}$\tablenotemark{b}}                                
& \colhead{}                                                                     
& \colhead{}                                                                     
& \colhead{}                                                                     
& \colhead{}                                                                     
& \colhead{}                                                                     
& \colhead{}                                                                     
& \colhead{}                                                                     
& \colhead{} \\                                                                  
\colhead{Galaxy}                                                                 
& \colhead{Model $Z$}                                                            
& \colhead{(yr)}                                                                 
& \colhead{(yr)}                                                                 
& \colhead{$\sigma$}                                                             
& \colhead{(M\protect\solar)}                                                    
& \colhead{$\sigma$}                                                             
& \colhead{(M\protect\solar/yr)}                                                 
& \colhead{$\sigma$}                                                             
& \colhead{N$_{mod}$\tablenotemark{c}}                                           
& \colhead{rms$_{mod}$\tablenotemark{d}}                                         
& \colhead{$\Delta(FUV-NUV)$\tablenotemark{e}}                                   
& \colhead{$\Delta(U-B)$\tablenotemark{e}}                                       
& \colhead{$\Delta(B-V)$\tablenotemark{e}}                                       
& \colhead{$\Delta(J-H)$\tablenotemark{e}}                                       
& \colhead{$\Delta(H-K)$\tablenotemark{e}}                                       
}                                                                                
\startdata                                                                       
\cutinhead{Im Galaxies}                                                          
DDO 50  \dotfill & 0.0040 &  7.00 &  8.12 & 0.07 &  7.78 & 0.03 & -0.04 & 0.05 & 0 & 0.15 & -0.02 &  0.19 & -0.09 & \nodata & \nodata \\
DDO 52  \dotfill & 0.0040 &  9.00 &  8.75 & 1.04 &  7.11 & 0.71 & -1.34 & 0.05 & 0 & 0.35 & -0.40 & -0.21 & -0.07 & \nodata & \nodata \\
DDO 53  \dotfill & 0.0040 &  0.00 &  9.47 & 0.69 &  7.15 & 0.40 & -2.02 & 0.30 & 6 & 0.45 & -0.27 &  0.30 & -0.10 &  0.03 & \nodata \\
DDO 63  \dotfill & 0.0004 &  8.00 &  8.69 & 0.24 &  7.31 & 0.16 & -1.07 & 0.05 & 0 & 0.05 & -0.04 &  0.00 & -0.02 & \nodata & \nodata \\
DDO 68  \dotfill & 0.0004 &  8.85 &  8.73 & 0.79 &  7.44 & 0.42 & -0.99 & 0.60 & 3 & 0.25 & -0.20 & -0.11 &  0.04 & \nodata & \nodata \\
DDO 70  \dotfill & 0.0004 &  8.48 &  9.16 & 0.05 &  7.34 & 0.05 & -1.52 & 0.05 & 0 & 0.05 &  0.01 & -0.02 &  0.03 & \nodata & \nodata \\
DDO 75  \dotfill & 0.0004 &  8.00 &  8.72 & 0.06 &  7.01 & 0.05 & -1.41 & 0.25 & 2 & 0.35 & -0.01 &  0.39 & -0.01 & \nodata & \nodata \\
DDO 99  \dotfill & 0.0040 &  8.00 &  8.74 & 0.07 &  7.42 & 0.06 & -1.02 & 0.05 & 0 & 0.15 & -0.04 &  0.17 &  0.02 & \nodata & \nodata \\
\enddata                                                                         
\tablenotetext{a}{A decay time of $0.00$ denotes a constant star formation rate.}
\tablenotetext{b}{Model mass times 2, divided by model age. The model mass is corrected for the mass that
has been returned to the interstellar medium.}
\tablenotetext{c}{Number of models with the same fit rms out of 7 models:        
constant SFR and 6 models with decaying SFRs.}                                   
\tablenotetext{d}{Overall rms of the model fit to the colors.}                   
\tablenotetext{e}{Model color minus the observed integrated color of the galaxy.}
\end{deluxetable}                                                                

\end{document}